\documentclass[11pt,a4paper]{article}
\usepackage{jheppub}
\usepackage[T1]{fontenc}
\usepackage{mathtools}
\usepackage{booktabs,array}
\usepackage{braket}
\usepackage{microtype}
\usepackage{cleveref}

\newcommand{\dd}{\mathrm d}
\newcommand{\e}{\mathrm e}
\newcommand{\cM}{\mathcal M}

\newcommand{\Hplus}{\mathcal H^{+}}

\newcommand{\cS}{\mathcal S}
\newcommand{\eps}{\varepsilon}

\newcommand{\sphere}{S^{D-2}}
\newcommand{\rh}{r_{\mathrm H}}
\newcommand{\Ehor}{\mathcal E}

\title{Soft-photon theorem near static spherical horizons in generic dimensions}

\author{Kai-Yu Zhang}
\affiliation{Institute for Theoretical Physics, TU Wien, Wiedner Hauptstrasse 8--10/136, A-1040 Vienna, Austria}
\emailAdd{kaiyu.zhang@student.tuwien.ac.at}

\abstract{We study Maxwell theory near a nonextremal bifurcate Killing horizon in a general static, spherically symmetric spacetime of dimension $D>3$. We impose near-horizon boundary conditions in a regular Eddington--Finkelstein chart that retain both radiative and Coulombic data, and construct the associated horizon phase space and large-$U(1)$ charge. At null infinity, the radiative falloff depends on the spacetime dimension and involves half-integer powers in odd $D$, whereas a nonextremal horizon at finite radius supports radiative data with an ordinary integer-power expansion. Expanding the horizon fields in Killing frequency and spherical harmonics, we derive the associated soft-photon relation from the large-$U(1)$ Ward identity. For comparison, we express the flat-space leading soft-photon theorem using the same spherical harmonics and commutator normalization as the horizon modes. In $D=4$, the two relations have the same soft-frequency dependence and the same angular dependence of the hard-charge terms, whereas for $D>4$ the flat-space relation contains additional dimension-dependent frequency and angular factors.}

\begin{document}

\maketitle

\section{Introduction}

Soft theorems are a central part of the infrared structure of gauge theory and gravity \cite{Weinberg:1965nx}. Their interplay with asymptotic symmetries and memory effects has been studied extensively in gravity, QED, and Yang--Mills theory \cite{Strominger:2013jfa,He:2014laa,He:2014cra,Strominger:2014pwa,Pasterski:2015zua,Strominger:2013lka,Campiglia:2015qka,Kapec:2015ena,He:2015zea,Mao:2017axa,Pate:2017vwa,Strominger:2017zoo}, and has also played an important role in the development of flat-space holography \cite{Pasterski:2016qvg,Pasterski:2017kqt,Raclariu:2021zjz,Donnay:2023mrd}. The soft expansion beyond leading order has likewise been studied in detail, including subleading soft theorems and, in gravity, the sub-subleading soft-graviton theorem \cite{Low:1958sn,Burnett:1967km,Lysov:2014csa,Casali:2014xpa,Campiglia:2014yka,Zlotnikov:2014sva,Campiglia:2016jdj,Laddha:2017ygw,Conde:2016csj,Conde:2016rom}. The relation between soft theorems and asymptotic charges also motivates the study of soft relations derived from horizon symmetries and charges. The soft-hair proposal of Hawking, Perry, and Strominger suggested that black holes carry soft hair on their horizons, associated with an infinite set of horizon charges, with possible implications for the black-hole information paradox \cite{Hawking:2016msc,Hawking:2016sgy}. This proposal led to extensive studies of near-horizon symmetries and charges \cite{Donnay:2015abr,Donnay:2016ejv,Shi:2016jtn,Grumiller:2019fmp,Adami:2020amw,Liu:2022uox}, electromagnetic and gravitational soft hair \cite{Averin:2016ybl,Mao:2016pwq,Choi:2018oel,Choi:2019fuq}, and horizon memory \cite{Donnay:2018ckb,Rahman:2019bmk}. The associated Ward identities provide the symmetry formulation of near-horizon soft-photon, soft-gluon, and soft-graviton theorems \cite{Cheng:2022xyr,Cheng:2022xgm,Mao:2023rca,Mao:2023dsy,Mao:2024phf,Zhuge:2026oap}. In particular, soft-graviton theorems have been derived from near-horizon supertranslation Ward identities for linearized gravity on Schwarzschild \cite{Mao:2023dsy} and de Sitter \cite{Mao:2024phf} backgrounds.

Soft theorems have been extended to higher spacetime dimensions in both gauge theory and gravity \cite{Schwab:2014xua,Zlotnikov:2014sva,Kalousios:2014uva,Kapec:2014zla,He:2019jjk,He:2019pll,He:2023bvv}. At null infinity, however, the form of the asymptotic expansion depends on the spacetime dimension. For $D>4$, the radiative and Coulombic sectors enter at different orders, and the structure of the expansion depends on whether $D$ is even or odd. In even dimensions the leading radiative falloff occurs at an integer power of $1/r$, whereas in odd dimensions it occurs at a half-integer power; the Coulombic sector remains at integer orders \cite{Hollands:2003ie,Hollands:2004ac,Tanabe:2011es,Campoleoni:2017qot,Satishchandran:2019pyc}. The even- and odd-dimensional cases must therefore be treated differently at null infinity. 

These dimension-dependent features at null infinity motivate a corresponding analysis near a smooth horizon at finite radius. We examine whether regular near-horizon boundary conditions retain both radiative and Coulombic data in every dimension $D>3$.

A comparison with the flat-space soft-photon theorem also requires a choice of photon modes and their normalization. The usual momentum-space formulation uses plane-wave states, while on a curved background the mode description depends on the available symmetries and the boundary conditions. For a horizon in a static, spherically symmetric spacetime, Killing frequency and spherical harmonics provide mode labels adapted to the geometry. A soft relation written in these modes can be compared with the flat-space theorem by expressing the latter in the same angular basis and keeping track of the mode normalization.

In this work we study Maxwell theory near a nonextremal bifurcate Killing horizon in a general static, spherically symmetric spacetime of dimension $D>3$. Working in a regular Eddington--Finkelstein chart, we impose a near-horizon falloff that retains both radiative and Coulombic data and solve the Maxwell equations as a characteristic hierarchy in $\rho=r-\rh$. We find that the smooth near-horizon expansion supports nonvanishing Maxwell news and finite radiative flux density in every dimension $D>3$, with no even--odd distinction in its radial power structure. The leading data also contain a Coulombic charge aspect, whose evolution is governed by a horizon Gauss law. The covariant symplectic form then defines the horizon phase space and the associated large-$U(1)$ charge, which separates into soft and hard contributions.

We construct regular near-horizon Maxwell modes with Killing frequency and spherical-harmonic labels and fix their normalization from the horizon symplectic form. The large-$U(1)$ charge selects the scalar-harmonic sector. After matching the symmetry across the bifurcation sphere, its Ward identity gives a soft-photon relation between the past and future horizon branches. We then express the flat-space leading soft-photon theorem using the same spherical harmonics and commutator normalization. This allows us to compare the frequency dependence and angular coefficients of the two soft relations. In $D=4$, they have the same soft-frequency dependence and the same angular dependence of the hard-charge terms. For $D>4$, the flat-space relation contains an additional dimension-dependent frequency weight and an $\ell$-dependent factor. This comparison does not identify the horizon and null-infinity scattering problems.

The paper is organized as follows. In section~\ref{sec:falloffsolutionspace}, we formulate the regular near-horizon falloffs for Maxwell theory, solve the resulting characteristic equations, and construct the horizon phase space and large-$U(1)$ charge. In section~\ref{sec:falloffs}, we compare the radiative expansions at null infinity and at the horizon, highlighting the even--odd distinction at null infinity and its absence for smooth near-horizon fields. In section~\ref{sec:softtheorem}, we quantize the horizon radiative modes, derive the soft-photon theorem from the large-$U(1)$ Ward identity, and compare it with the flat-space leading soft-photon theorem in a common spherical-harmonic basis. We conclude with a discussion of the main results and their implications. Appendix~\ref{app:flatsoft} gives the detailed derivation of the flat-space soft theorem in this basis.

\section{Regular near-horizon fall-off and solution space}
\label{sec:falloffsolutionspace}
In this section we analyze Maxwell theory near a nonextremal bifurcate Killing horizon in a general static, spherically symmetric spacetime of dimension $D>3$.\footnote{We restrict to $D>3$. In three dimensions, QED has a logarithmic Coulomb interaction and a distinct infrared sector \cite{Bosma:2023sxn,Shimizu:2025hfl}; gravity also exhibits soft-graviton relations, BMS$_3$ symmetry, and memory despite having no propagating bulk gravitons \cite{Cotler:2024cia}. These cases lie outside our scope.} We impose a regular near-horizon fall-off that retains both finite radiative data and a regular Coulombic sector, solve the resulting characteristic hierarchy, and construct the associated horizon phase space. The residual large-$U(1)$ symmetry is generated by a charge that separates into soft and hard contributions.

\subsection{Near-horizon $U(1)$ fall-off}

We consider Maxwell theory near a smooth, nonextremal bifurcate Killing horizon in a fixed static, spherically symmetric spacetime of dimension $D>3$. The future and past horizons, $\Hplus$ and $\mathcal H^-$, intersect at a bifurcation sphere $B\simeq S^{D-2}$. Staticity is imposed as a separate assumption on the background, rather than inferred from spherical symmetry \cite{Chrusciel:2012jk}. Examples include the nonextremal horizons of Schwarzschild--Tangherlini and Reissner--Nordstr\"om black holes, their asymptotically AdS counterparts, and de Sitter cosmological horizons.

\begin{figure}[t]
\centering
\includegraphics[width=0.72\textwidth]{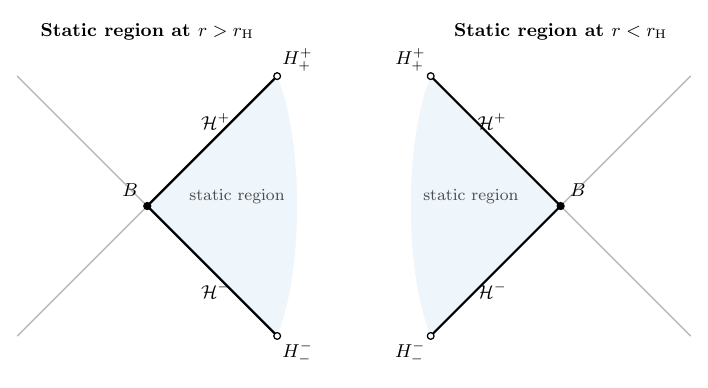}
\caption{Local causal structure of the two nonextremal horizon orientations. The shaded region is static, lying at $r>\rh$ on the left and $r<\rh$ on the right. The diagrams show only the local horizon structure and do not specify the asymptotic boundary.}
\label{fig:horizon-orientations}
\end{figure}

We use Eddington--Finkelstein coordinates $x^\mu=(v,r,x^A)$ adapted to $\Hplus$, with $r$ the areal radius and $\xi=\partial_v$ future-directed. The metric takes the form
\begin{equation}
\dd s^2=-F(r) \dd v^2+2H(r) \dd v \dd r+r^2q_{AB}(x) \dd x^A\dd x^B,
\label{eq:metric}
\end{equation}
where $q_{AB}$ is the metric on the unit sphere $S^{D-2}$. The sign of $H$ depends on which side of the horizon contains the static region. For a Schwarzschild-type future horizon, the static region lies at $r>\rh$ and $v$ is an advanced coordinate, with $H>0$. For the future cosmological horizon of a de Sitter static patch, the static region lies at $r<\rh$ and $v$ is a retarded coordinate, with $H<0$. We use the same notation $v$ in both cases and orient horizon integrals accordingly.  The two possible local orientations of the static region are illustrated in figure~\ref{fig:horizon-orientations}.
Related near-horizon soft theorems have been studied for gauge fields in Schwarzschild and de Sitter spacetimes \cite{Cheng:2022xyr,Mao:2023rca}, and for linearized gravity on these backgrounds \cite{Mao:2023dsy,Mao:2024phf}.

The inverse metric and volume density are
\begin{equation}
\begin{gathered}
g^{vv}=0,\qquad g^{vr}=\frac{1}{H},\qquad g^{rr}=\frac{F}{H^2},\qquad g^{AB}=\frac{q^{AB}}{r^2},\qquad \sqrt{-g}=|H|r^{D-2}\sqrt q.
\end{gathered}
\end{equation}
The nonzero Christoffel symbols are
\begin{equation}
\begin{aligned}
\Gamma^v{}_{vv}&=\frac{F'}{2H}, & \Gamma^v{}_{AB}&=-\frac{r}{H}q_{AB}, & \Gamma^r{}_{vv}&=\frac{FF'}{2H^2}, & \Gamma^r{}_{vr}&=-\frac{F'}{2H}, \\
\Gamma^r{}_{rr}&=\frac{H'}{H}, & \Gamma^r{}_{AB}&=-\frac{rF}{H^2}q_{AB}, & \Gamma^A{}_{rB}&=\frac{1}{r}\delta^A{}_B, & \Gamma^A{}_{BC}&=\Gamma^A{}_{BC}[q].
\end{aligned}
\end{equation}
Here a prime denotes differentiation with respect to $r$.

The horizon is located at $r=\rh>0$, where regularity and nonextremality require
\begin{equation}
F(\rh)=0,\qquad H_{\mathrm H}:=H(\rh)\ne0,\qquad F'(\rh)\ne0.
\end{equation}
The surface gravity associated with $\xi=\partial_v$ follows from $\xi^\nu\nabla_\nu\xi^\mu=\kappa\xi^\mu$ on $\Hplus$:
\begin{equation}
\kappa=\frac{F'(\rh)}{2H_{\mathrm H}}.
\end{equation}
With the future-horizon conventions above, $\kappa>0$ in both cases: $F'(\rh)$ and $H_{\mathrm H}$ are positive for the black-hole horizon and negative for the de Sitter cosmological horizon.

Defining $\rho:=r-\rh$ and $K(r):=F(r)/H(r)$, the near-horizon expansions are
\begin{equation}
\begin{aligned}
F(r)=2\kappa H_{\mathrm H}\rho+O(\rho^2),\quad H(r)=H_{\mathrm H}+H'(\rh)\rho+O(\rho^2),\quad K(r)&=2\kappa\rho+O(\rho^2).
\end{aligned}
\end{equation}
The chart is regular on $\Hplus$ at finite $v$. Writing $\dd r_*/\dd r=H/F$, the static time is $t=v-r_*$, and the coordinate $u=t-r_*$ gives the corresponding chart on $\mathcal H^-$,
\begin{equation}
\dd s^2=-F(r) \dd u^2-2H(r) \dd u \dd r+r^2q_{AB} \dd x^A\dd x^B.
\end{equation}
The static time reflection $t\mapsto-t$ extends across $B$ and interchanges the two horizon branches, sending $v$ to $-u$. We work locally on $\Hplus$ and use the bifurcate extension to match the past and future horizon data.

We first treat matter as a prescribed conserved current $J^\mu$ coupled to the Maxwell field,
\begin{equation}
S[A;J]=-\frac{1}{4g_D^2}\int_{\cM}\dd^Dx\sqrt{-g} F_{\mu\nu}F^{\mu\nu}-\int_{\cM}\dd^Dx\sqrt{-g} A_\mu J^\mu,\qquad F=\dd A.
\end{equation}
The equations of motion and gauge transformation are
\begin{equation}
\nabla_\mu F^{\mu\nu}=g_D^2J^\nu,\qquad \nabla_\mu J^\mu=0,\qquad \delta_\eps A_\mu=\partial_\mu\eps.
\label{eq:Maxwell}
\end{equation}

We restrict to Maxwell fields that are smooth in the regular Eddington--Finkelstein chart. In radial gauge, we further impose $A_v|_{\Hplus}=0$ and adopt the near-horizon conditions
\begin{equation}
A_r=0,\qquad A_v=O(\rho),\qquad A_A=O(1).
\label{eq:falloff}
\end{equation}
These conditions retain time-dependent angular data $a_A(v,x)=A_A|_{\Hplus}$ and a finite, potentially nonzero electric component $e(v,x)=F_{rv}|_{\Hplus}$. They therefore allow both radiative and Coulombic data.

Preserving $A_r=0$ gives $\partial_r\eps=0$, while preserving $A_v=O(\rho)$ further requires $\partial_v\eps=0$. Hence
\begin{equation}
\eps=\eps(x^A).
\label{eq:lgauge}
\end{equation}
For smooth near-horizon fields, these falloffs also give a finite horizon energy-flux density,
\begin{equation}
T^{\mathrm M}_{vv}=\frac{1}{g_D^2}\left[\frac{F}{H^2}(F_{vr})^2+\frac{1}{r^2}F_{vA}F_v{}^A+\frac{F}{4}F_{\alpha\beta}F^{\alpha\beta}\right]\longrightarrow\frac{1}{g_D^2\rh^2}F_{vA}F_v{}^A\big|_{\Hplus}.
\label{eq:flux}
\end{equation}
Since $F_{\alpha\beta}F^{\alpha\beta}$ is finite and $H_{\mathrm H}\ne0$, the terms proportional to $F$ vanish on $\Hplus$. 

The corresponding near-horizon expansion is
\begin{align}
A_v&=\rho e(v,x)+\frac{\rho^2}{2}e_{(1)}(v,x)+O(\rho^3), \\
A_A&=a_A(v,x)+\rho b_A(v,x)+\frac{\rho^2}{2}b_{(1)A}(v,x)+O(\rho^3), \\
J^\mu&=J_{(0)}^\mu(v,x)+\rho J_{(1)}^\mu(v,x)+O(\rho^2).
\end{align}
A finite nonzero $J^r_{(0)}=J^r|_{\Hplus}$ is allowed and provides the matter flux entering the hard charge.

To determine the leading horizon data, we expand the field strength to first subleading order:
\begin{align}
F_{rv}&=e+\rho e_{(1)}+O(\rho^2),\\
F_{rA}&=b_A+\rho b_{(1)A}+O(\rho^2),\\
F_{vA}&=\partial_va_A+\rho\bigl(\partial_vb_A-D_Ae\bigr)+O(\rho^2),\\
F_{AB}&=D_Aa_B-D_Ba_A+\rho\bigl(D_Ab_B-D_Bb_A\bigr)+O(\rho^2).
\end{align}

\subsection{Solution space}

With the gauge choice and falloffs \eqref{eq:falloff}, the Maxwell equations \eqref{eq:Maxwell} form a characteristic hierarchy. Current conservation reads
\begin{equation}
\partial_v\left(Hr^{D-2}J^v\right)+\partial_r\left(Hr^{D-2}J^r\right)+Hr^{D-2}D_AJ^A=0.
\label{eq:Jcons}
\end{equation}
The $v$ component of the Maxwell equations is the hypersurface equation,
\begin{equation}
-\partial_r\left(\frac{r^{D-2}}{H}F_{rv}\right)-r^{D-4}D_AF_r{}^A=g_D^2Hr^{D-2}J^v,
\label{eq:Mv}
\end{equation}
while the angular components give the evolution equations,
\begin{equation}
\partial_v\left(r^{D-4}F_{rA}\right)+\partial_r\left[r^{D-4}\left(F_{vA}+KF_{rA}\right)\right]+Hr^{D-6}D^BF_{BA}=g_D^2Hr^{D-2}q_{AB}J^B.
\label{eq:MA}
\end{equation}
Here and below angular indices are raised and lowered with $q_{AB}$, in particular $F_v{}^A=q^{AB}F_{vB}$ and $F_r{}^A=q^{AB}F_{rB}$. The remaining $r$ component is the supplementary equation,
\begin{equation}
\partial_v\left(\frac{r^{D-2}}{H}F_{rv}\right)-r^{D-4}D_A\left(F_v{}^A+KF_r{}^A\right)=g_D^2Hr^{D-2}J^r.
\label{eq:Mr}
\end{equation}

Current conservation and the antisymmetry of $F^{\mu\nu}$ imply the Noether identity
\begin{equation}
\nabla_\nu\left[g_D^2J^\nu-\nabla_\mu F^{\mu\nu}\right]=0.
\end{equation}
Once \eqref{eq:Jcons}, \eqref{eq:Mv}, and \eqref{eq:MA} are satisfied, this reduces to
\begin{equation}
\partial_r\left(\sqrt{-g}\left(g_D^2J^r-\nabla_\mu F^{\mu r}\right)\right)=0.
\label{eq:supp}
\end{equation}
The supplementary equation therefore needs to be imposed at only one radial order. Since $F=\dd A$, the Bianchi identity holds identically. Its $(r,v,A)$ component,
\begin{equation}
\partial_rF_{vA}=\partial_vF_{rA}-D_AF_{rv},
\label{eq:Bianchi}
\end{equation}
will be used in the angular equation to eliminate $\partial_rF_{vA}$.

The hypersurface equation \eqref{eq:Mv} at $O(1)$ gives
\begin{equation}
e_{(1)}=-\left(\frac{D-2}{\rh}-\frac{H'(\rh)}{H_{\mathrm H}}\right)e-\frac{H_{\mathrm H}}{\rh^2}D_Ab^A-g_D^2H_{\mathrm H}^2J^v_{(0)}.
\end{equation}
Thus $e_{(1)}$ is fixed algebraically by the leading horizon data.

The leading angular equation follows from \eqref{eq:MA}. Using \eqref{eq:Bianchi} and $K=2\kappa\rho+O(\rho^2)$ gives
\begin{equation}
\left(2\partial_v+2\kappa\right)b_A=D_Ae-\frac{D-4}{\rh}\partial_va_A-\frac{H_{\mathrm H}}{\rh^2}D^B\left(D_Ba_A-D_Aa_B\right)+g_D^2H_{\mathrm H}r_{\rm H}^2q_{AB}J^{B}_{(0)}.
\end{equation}
This determines the evolution of the first radial coefficient $b_A$ along the horizon.

Finally, the $O(1)$ part of the supplementary equation \eqref{eq:Mr} gives
\begin{equation}
\partial_ve=\frac{H_{\mathrm H}}{\rh^2}D_A\partial_va^A+g_D^2H_{\mathrm H}^2J^r_{(0)}.
\end{equation}
All higher radial orders of the supplementary equation follow from \eqref{eq:supp}. Thus $a_A(v,x)$ is the free radiative datum, and the remaining radial coefficients are determined recursively once the transverse initial data are specified. We define the gauge-invariant Maxwell \textit{news} by
\begin{equation}
N_A:=F_{vA}\big|_{\Hplus}=\partial_va_A,
\end{equation}
with $D-2$ local components. We define the electric charge aspect by
\begin{equation}
\Ehor:=\frac{\rh^{D-2}}{H_{\mathrm H}}e.
\label{eq:Ehor}
\end{equation}
It obeys the horizon Gauss law,
\begin{equation}
\partial_v\Ehor=\rh^{D-4}D^AN_A+g_D^2H_{\mathrm H}\rh^{D-2}J^r\big|_{\Hplus}.
\label{eq:Gauss}
\end{equation}

The regular near-horizon falloffs thus admit Maxwell radiation. At null infinity, a finite nonzero radiative flux requires the radiative potential to fall off as $r^{-(D-2)/2}$ in Cartesian components \cite{Campoleoni:2017qot,He:2019jjk}. This gives half-integer powers in odd spacetime dimensions. We compare the two expansions in section~\ref{sec:falloffs}.

\subsection{Horizon phase space and large-$U(1)$ charge}
\label{sec:phase}

The two ends of $\Hplus$ play different roles. The limit $v\to-\infty$ approaches the bifurcation sphere $B$, while $v\to+\infty$ is the future endpoint of the horizon.
Following the soft-photon phase-space construction of \cite{Strominger:2017zoo,He:2014cra,He:2019jjk,Strominger:2013lka}, we impose vanishing magnetic field at both ends and vanishing electric charge aspect at the future endpoint,
\begin{equation}
F_{AB}\big|_{v\to\pm\infty}(\rh,x)=0,\qquad \Ehor\big|_{v\to+\infty}(x)=0.
\end{equation}
In the topologically trivial magnetic sector, the integrated news is therefore an exact one-form,
\begin{equation}
\Delta a_A:=\int_{-\infty}^{+\infty}\dd v N_A=D_A\mathcal N.
\end{equation}

At the bifurcation sphere we impose the horizon matching condition
\begin{equation}
\lim_{u\to+\infty}\frac{F_{ru}^{(-)}(u,\rh,x)}{H_{\mathrm H}}=\lim_{v\to-\infty}\frac{F_{rv}^{(+)}(v,\rh,x)}{H_{\mathrm H}},\qquad \eps^{-}(x)=\eps^{+}(x)\equiv\eps(x),
\label{eq:Hmatch}
\end{equation}
where the angular coordinates are identified on $B$ and the same Killing-time normalization is used on the two branches.

The Maxwell symplectic potential is
\begin{equation}
\Theta_{\mathrm M}(\delta A)=-\frac{1}{g_D^2} \delta A\wedge *F.
\end{equation}
Using $A_v|_{\Hplus}=\delta A_v|_{\Hplus}=0$ and $\sqrt{-g}F^{rA}|_{\Hplus}=\operatorname{sgn}(H_{\mathrm H})\rh^{D-4}\sqrt q N^A$, we integrate the associated symplectic current over the horizon to obtain
\begin{equation}
\Omega_{\Hplus}^{\mathrm M}=\frac{\rh^{D-4}}{g_D^2}\int\dd v \dd\Omega_{D-2}\left(\delta_1N^A \delta_2a_A-\delta_2N^A \delta_1a_A\right).
\label{eq:OmegaM}
\end{equation}

With $N_A=\partial_va_A$, the symplectic operator in \eqref{eq:OmegaM} is proportional to $\partial_v$ and has a $v$-independent zero mode. Removing this mode, we obtain the radiative bracket
\begin{equation}
\{N_A(v,x),a_B(v',x')\}_{\mathrm{rad}}=-\frac{g_D^2}{2\rh^{D-4}} q_{AB}(x) \delta(v-v') \delta_{D-2}(x,x'),
\label{eq:PB}
\end{equation}
where $\delta_{D-2}(x,x')$ is normalized with respect to $\dd\Omega_{D-2}$. The factor $1/2$ follows because the two antisymmetrized terms in \eqref{eq:OmegaM} contribute equally after integration by parts.

The residual gauge transformation \eqref{eq:lgauge} acts on the leading angular field as
\begin{equation}
\delta_\eps a_A=D_A\eps.
\label{eq:U1A}
\end{equation}
In terms of \eqref{eq:Ehor}, the charge on a constant-$v$ cut of $\Hplus$ is
\begin{equation}
Q_\eps(v)=-\frac{1}{g_D^2}\int_{\sphere}\dd\Omega_{D-2} \eps \Ehor(v,x)=-\frac{\rh^{D-2}}{g_D^2H_{\mathrm H}}\int_{\sphere}\dd\Omega_{D-2} \eps F_{rv}\big|_{\Hplus}.
\end{equation}
The charge entering the horizon matching is its value on the bifurcation sphere,
\begin{equation}
Q_\eps^{B}:=\lim_{v\to-\infty}Q_\eps(v)=-\frac{1}{g_D^2}\int_{\sphere}\dd\Omega_{D-2} \eps \Ehor^{B}(x).
\end{equation}
The matching condition \eqref{eq:Hmatch} identifies this with the charge on $\mathcal H^-$ at $B$. Integrating the Gauss law \eqref{eq:Gauss} with the endpoint condition above gives
\begin{equation}
-Q_\eps^{B}=Q_\eps^{\mathrm S}+Q_\eps^{\mathrm H},
\end{equation}
where
\begin{align}
Q_\eps^{\mathrm S}&=\frac{\rh^{D-4}}{g_D^2}\int_{-\infty}^{\infty}\dd v\int_{\sphere}\dd\Omega_{D-2} D^A\eps N_A =-\frac{\rh^{D-4}}{g_D^2}\int_{\sphere}\dd\Omega_{D-2} \eps D^2\mathcal N, \label{eq:Qsoft}\\
Q_\eps^{\mathrm H}&=-H_{\mathrm H}\rh^{D-2}\int_{-\infty}^{\infty}\dd v\int_{\sphere}\dd\Omega_{D-2} \eps J^r\big|_{\Hplus}.\label{eq:Qhard}
\end{align}

Using only the bulk radiative bracket \eqref{eq:PB}, the soft charge gives
\begin{equation}
\{a_A(v,x),Q_\eps^{\mathrm S}\}_{\mathrm{rad}}=\frac12D_A\eps(x).
\end{equation}
The missing contribution comes from the endpoint zero mode, as at null infinity \cite{Strominger:2013lka,Strominger:2017zoo,He:2020ifr}. Write $a_A^\pm=D_A\varphi_\pm$, $\varphi=(\varphi_++\varphi_-)/2$, $\mathcal N=\varphi_+-\varphi_-$, and $a_A=\bar a_A+D_A\varphi$, so that $\bar a_A^\pm=\pm D_A\mathcal N/2$. Keeping the endpoint variations in \eqref{eq:OmegaM} gives
\begin{equation}
\Omega_{\mathrm{ext}}^{\mathrm M}=\frac{\rh^{D-4}}{g_D^2}\left[\int\dd v \dd\Omega_{D-2} \delta\partial_v\bar a^A\wedge\delta\bar a_A+\int\dd\Omega_{D-2} D^A\delta\mathcal N\wedge D_A\delta\varphi\right].
\end{equation}
Its contraction with $\delta_\eps\varphi=\eps$, $\delta_\eps\bar a_A=0$ gives $\delta Q_\eps^{\mathrm S}$. The bracket on the extended phase space therefore satisfies
\begin{equation}
\{a_A(v,x),Q_\eps^{\mathrm S}\}=D_A\eps(x),
\end{equation}
in agreement with \eqref{eq:U1A}.

For the hard sector, take a massless complex scalar $\Phi$ of charge $q$, with $\mathcal D_\mu\Phi=(\nabla_\mu-iqA_\mu)\Phi$ and action
\begin{equation}
S_\Phi[A,\Phi]=-\int_{\cM}\dd^Dx\sqrt{-g} (\mathcal D_\mu\Phi)^*\mathcal D^\mu\Phi.
\end{equation}
The corresponding current is
\begin{equation}
J^\mu=iq\left[\Phi^*\mathcal D^\mu\Phi-(\mathcal D^\mu\Phi)^*\Phi\right].
\end{equation}
For $\Phi=\phi(v,x)+O(\rho)$ near $\Hplus$, the horizon current is $H_{\mathrm H}J^r=iq(\phi^*\partial_v\phi-\partial_v\phi^* \phi)$. Using $\mathcal D^r\Phi=H_{\mathrm H}^{-1}\partial_v\phi$ and the chosen horizon orientation, the symplectic form reduces to
\begin{equation}
\Omega_{\Hplus}^{\Phi}=-\rh^{D-2}\int\dd v \dd\Omega_{D-2}\left[\delta_1\phi^* \partial_v\delta_2\phi+\delta_1\phi \partial_v\delta_2\phi^*-(1\leftrightarrow2)\right].
\end{equation}
Together with \eqref{eq:Qhard}, this gives
\begin{equation}
\{\phi(v,x),Q_\eps^{\mathrm H}\}=iq\eps(x)\phi(v,x),\qquad \{\phi^*(v,x),Q_\eps^{\mathrm H}\}=-iq\eps(x)\phi^*(v,x).
\label{eq:hard-act}
\end{equation}
Thus $Q_\eps^{\mathrm S}$ and $Q_\eps^{\mathrm H}$ generate the large-$U(1)$ transformations of the Maxwell and charged-matter sectors, respectively.

\section{Radiative falloffs at null infinity and the horizon}
\label{sec:falloffs}
We compare the radial behavior of radiative Maxwell fields at null infinity with that near a smooth nonextremal horizon. The asymptotic structure of higher-dimensional gauge fields has been studied extensively from the viewpoints of large gauge symmetries, soft theorems, memory, and asymptotic phase space \cite{Kapec:2014zla,Campoleoni:2017qot,He:2019jjk,He:2019pll,Henneaux:2019yqq,Campoleoni:2019ptc,Esmaeili:2019hom,Freidel:2019ohg}. At null infinity, the radiative falloff depends on the spacetime dimension; in odd dimensions the radiative branch involves half-integer powers of $1/r$, whereas the Coulombic branch is organized in integer powers \cite{Campoleoni:2017qot,Satishchandran:2019pyc,Henneaux:2019yqq}. Near the horizon, by contrast, smooth radiative data admit a regular expansion in integer powers of $\rho$.

\subsection{Radiative falloffs at null infinity}

Null infinity lies at $r\to\infty$, where outgoing solutions determine the radial falloff. In odd spacetime dimensions the retarded Green function of the massless wave equation has support inside the light cone and produces tails \cite{Satishchandran:2017pek}. The Maxwell radiative branch then has half-integer powers of $1/r$, while the Coulombic branch has integer powers \cite{Henneaux:2019yqq}. We derive the radiative falloff from the retarded Green function.

Consider free Maxwell theory in $D$-dimensional Minkowski spacetime, with Cartesian coordinates $X^\alpha=(t,x^i)$, $i=1,\ldots,D-1$. In Lorenz gauge, $\partial^\alpha A_\alpha=0$, the source-free equations reduce to
\begin{equation}
\left(-\partial_t^2+\delta^{ij}\partial_i\partial_j\right)A_\alpha=0.
\end{equation}
Let $G_{\mathrm{ret}}$ satisfy
\begin{equation}
\left(-\partial_t^2+\delta^{ij}\partial_i\partial_j\right)G_{\mathrm{ret}}(X-X')=-\delta^{(D)}(X-X').
\end{equation}
On the Cauchy surface $\Sigma_0:t=t_0$, define
\begin{equation}
A_\alpha^{(0)}(\mathbf{x}):=A_\alpha(t_0,\mathbf{x}),\qquad \Pi_\alpha(\mathbf{x}):=\partial_tA_\alpha(t_0,\mathbf{x}).
\end{equation}
The solution to the future of $\Sigma_0$ is
\begin{equation}
A_\alpha(t,\mathbf{x})=\int_{\Sigma_0}\dd^{D-1}x'\left[G_{\mathrm{ret}}(t-t_0,\mathbf{x}-\mathbf{x}') \Pi_\alpha(\mathbf{x}')-A_\alpha^{(0)}(\mathbf{x}') \partial_{t'}G_{\mathrm{ret}}(t-t',\mathbf{x}-\mathbf{x}')\big|_{t'=t_0}\right].
\label{eq:Green}
\end{equation}

Writing $R:=|\mathbf{x}-\mathbf{x}'|$, the retarded Green function may be decomposed as
\begin{equation}
G_{\mathrm{ret}}(t-t',R)=\int_0^\infty\frac{\dd\omega}{2\pi} \e^{-i\omega(t-t')}G_\omega(R)+\mathrm{c.c.},
\label{eq:Gfreq}
\end{equation}
where
\begin{equation}
\left(\nabla_{D-1}^2+\omega^2\right)G_\omega(\mathbf{x}-\mathbf{x}')=-\delta^{(D-1)}(\mathbf{x}-\mathbf{x}')
\end{equation}
and the outgoing solution is
\begin{equation}
G_\omega(R)=\frac{i}{4}\left(\frac{\omega}{2\pi R}\right)^{\frac{D-3}{2}}H^{(1)}_{\frac{D-3}{2}}(\omega R).
\end{equation}
Here $H_\nu^{(1)}$ is the Hankel function of the first kind of order $\nu$. For fixed nonzero $\omega$ and large $R$,
\begin{equation}
H_\nu^{(1)}(z)=\sqrt{\frac{2}{\pi z}} \e^{i(z-\pi\nu/2-\pi/4)}\left[1+O(z^{-1})\right],
\end{equation}
which gives
\begin{equation}
G_\omega(R)=\frac{i \e^{-i\pi(D-2)/4}}{2(2\pi)^{\frac{D-2}{2}}} \omega^{\frac{D-4}{2}}R^{-\frac{D-2}{2}}\e^{i\omega R}\left[1+O\left((\omega R)^{-1}\right)\right].
\label{eq:Gfar}
\end{equation}
The far-zone expansion of $R$ at $\mathbf{x}=r\hat{\mathbf{x}}$ is
\begin{equation}
R=|r\hat{\mathbf{x}}-\mathbf{x}'|=r-\hat{\mathbf{x}}\cdot\mathbf{x}'+O(r^{-1}).
\label{eq:Rfar}
\end{equation}
Substituting \eqref{eq:Gfreq}, \eqref{eq:Gfar}, and \eqref{eq:Rfar} into \eqref{eq:Green} gives
\begin{equation}
A_\alpha^{\mathrm{rad}}(t,r,\hat{\mathbf{x}})=r^{-\frac{D-2}{2}}\int_0^\infty\frac{\dd\omega}{2\pi} \e^{-i\omega(t-r)}\widetilde a_\alpha(\omega,\hat{\mathbf{x}})+\mathrm{c.c.}+O\left(r^{-\frac D2}\right),
\end{equation}
where
\begin{equation}
\widetilde a_\alpha(\omega,\hat{\mathbf{x}})=\frac{i \e^{-i\pi(D-2)/4}}{2(2\pi)^{\frac{D-2}{2}}} \omega^{\frac{D-4}{2}}\e^{i\omega t_0}\int\dd^{D-1}x' \e^{-i\omega\hat{\mathbf{x}}\cdot\mathbf{x}'}\left[\Pi_\alpha(\mathbf{x}')-i\omega A_\alpha^{(0)}(\mathbf{x}')\right].
\end{equation}
With $u=t-r$ and
\begin{equation}
a_\alpha(u,\hat{\mathbf{x}}):=\int_0^\infty\frac{\dd\omega}{2\pi} \widetilde a_\alpha(\omega,\hat{\mathbf{x}})\e^{-i\omega u}+\mathrm{c.c.},
\end{equation}
the radiative solution is
\begin{equation}
A_\alpha^{\mathrm{rad}}(t,r,\hat{\mathbf{x}})=r^{-\frac{D-2}{2}}a_\alpha(u,\hat{\mathbf{x}})+O\left(r^{-\frac D2}\right).
\label{eq:Arad}
\end{equation}

Introduce retarded coordinates $(u,r,x^A)$ through $x^i=r\hat x^i(x^A)$. Since
\begin{equation}
\dd t=\dd u+\dd r,\qquad \dd x^i=\hat x^i\dd r+r \partial_A\hat x^i\dd x^A,
\end{equation}
the potential components are
\begin{equation}
A_u=A_t,\qquad A_r=A_t+\hat x^iA_i,\qquad A_A=rA_i\partial_A\hat x^i.
\end{equation}
The leading Lorenz-gauge condition obtained from \eqref{eq:Arad} is
\begin{equation}
\partial_u\left(a_t+\hat x^ia_i\right)=0.
\end{equation}
For the nonzero-frequency radiative part,
\begin{equation}
a_t+\hat x^ia_i=0.
\label{eq:trans}
\end{equation}
Using \eqref{eq:trans}, the radiative potential in retarded coordinates has the falloff
\begin{align}
A_u^{\mathrm{rad}}&=r^{-\frac{D-2}{2}}a_t(u,x)+O\left(r^{-\frac D2}\right),& A_r^{\mathrm{rad}}&=O\left(r^{-\frac D2}\right),\\
A_A^{\mathrm{rad}}&=r^{-\frac{D-4}{2}}a_i(u,x)\partial_A\hat x^i+O\left(r^{-\frac{D-2}{2}}\right).
\end{align}

The same leading angular falloff is obtained in radial gauge. Since $A_r^{\mathrm{rad}}=O(r^{-D/2})$, the gauge parameter
\begin{equation}
\eps_{\mathrm{rg}}(u,r,x):=\int_r^\infty\dd r' A_r^{\mathrm{rad}}(u,r',x)=O\left(r^{-\frac{D-2}{2}}\right)
\end{equation}
sets $A_r+\partial_r\eps_{\mathrm{rg}}=0$. It changes $A_u$ and $A_A$ only by
\begin{equation}
\partial_u\eps_{\mathrm{rg}}=O\left(r^{-\frac{D-2}{2}}\right),\qquad \partial_A\eps_{\mathrm{rg}}=O\left(r^{-\frac{D-2}{2}}\right),
\end{equation}
so the leading $r^{-(D-4)/2}$ term in $A_A^{\mathrm{rad}}$ is unchanged.

The leading radiative field strength is therefore
\begin{equation}
F_{uA}^{\mathrm{rad}}=r^{-\frac{D-4}{2}}\bar{N}_A(u,x)+O\left(r^{-\frac{D-2}{2}}\right),\qquad \bar{N}_A:=\partial_u a_i \partial_A\hat x^i,
\end{equation}
while
\begin{equation}
F_{ur}^{\mathrm{rad}}=O\left(r^{-\frac D2}\right).
\end{equation}
Using $g^{AB}=r^{-2}q^{AB}$, the leading energy-flux density is
\begin{equation}
T_{uu}^{\mathrm M}=\frac{1}{g_D^2r^2}q^{AB}F_{uA}^{\mathrm{rad}}F_{uB}^{\mathrm{rad}}+O\left(r^{-(D-1)}\right)=\frac{1}{g_D^2}r^{-(D-2)}\bar{N}_A\bar{N}^A+O\left(r^{-(D-1)}\right).
\end{equation}

For comparison, consider the Coulomb field sourced by a static charge $Q$, with
\begin{equation}
Q:=\frac{1}{g_D^2}\int_{S_r^{D-2}}r^{D-2}\dd\Omega_{D-2} F_{ur}^{\mathrm C}.
\end{equation}
Spherical symmetry gives
\begin{equation}
F_{ur}^{\mathrm C}=\frac{g_D^2Q}{\Omega_{D-2}} r^{-(D-2)}.
\end{equation}
The Coulombic field has integer powers for every $D$, whereas $F_{uA}^{\mathrm{rad}}$ starts at order $r^{-(D-4)/2}$. An expansion restricted to integer powers of $1/r$ therefore misses the leading radiation in odd $D$ \cite{Campoleoni:2017qot,He:2019jjk}.

\subsection{Regular radiative falloffs at the horizon}
\label{sec:NHrad}

The horizon $\rho=0$ is a regular characteristic surface. We now construct the smooth radiative branch as an expansion in integer powers of $\rho$, setting $J^\mu=0$ and keeping radial gauge $A_r=0$.

The remaining components are expanded in scalar and transverse vector harmonics on $S^{D-2}$,
\begin{align}
A_v(v,r,x)&=\sum_{\ell=0}^{\infty}\sum_m\alpha_{\ell m}(v,r)Y_{\ell m}(x),\\
A_A(v,r,x)&=\sum_{\ell=1}^{\infty}\sum_m\gamma_{\ell m}(v,r)D_AY_{\ell m}(x)+\sum_{\ell=1}^{\infty}\sum_p\Psi^V_{\ell p}(v,r)V_A^{\ell p}(x),
\end{align}
with $\alpha_{\ell m}(v,\rh)=0$ from $A_v=O(\rho)$. The harmonics satisfy
\begin{equation}
-D^2Y_{\ell m}=\lambda_\ell^S Y_{\ell m},\qquad \lambda_\ell^S=\ell(\ell+D-3),
\label{eq:Ylm}
\end{equation}
and
\begin{equation}
D^AV_A^{\ell p}=0,\qquad D_B\left(D^BV_A^{\ell p}-D_AV^{B,\ell p}\right)=-\lambda_\ell^V V_A^{\ell p},\qquad \lambda_\ell^V=(\ell+1)(\ell+D-4).
\label{eq:Vlm}
\end{equation}

Since $D_AY_{00}=0$, the scalar monopole has no angular component. Equations \eqref{eq:Mv} and \eqref{eq:Mr} in vacuum reduce to
\begin{equation}
\partial_r\left(\frac{r^{D-2}}{H}F_{rv,00}^S\right)=0,\qquad \partial_v\left(\frac{r^{D-2}}{H}F_{rv,00}^S\right)=0.
\end{equation}
Hence
\begin{equation}
\frac{r^{D-2}}{H}F_{rv,00}^S=Q_0,
\end{equation}
with $Q_0$ constant. This is the Coulombic monopole.

For $\ell\ge1$, let $F_{rv,\ell m}^S$ denote the coefficient of $Y_{\ell m}$ in $F_{rv}$ and define
\begin{equation}
\Psi^S_{\ell m}:=\frac{r^{D-2}}{H}F_{rv,\ell m}^S.
\end{equation}
Equations \eqref{eq:Mv} and \eqref{eq:Mr} give the remaining scalar field-strength components,
\begin{align}
F_{rA}^{S,\ell m}&=\frac{r^{4-D}}{\lambda_\ell^S}\partial_r\Psi^S_{\ell m} D_AY_{\ell m},\\
F_{vA}^{S,\ell m}&=-\frac{r^{4-D}}{\lambda_\ell^S}\left(\partial_v\Psi^S_{\ell m}+K\partial_r\Psi^S_{\ell m}\right)D_AY_{\ell m}.
\label{eq:FS}
\end{align}
Substitution into the Bianchi identity \eqref{eq:Bianchi} gives
\begin{equation}
2\partial_v\partial_r\Psi^S_{\ell m}+\frac{4-D}{r}\partial_v\Psi^S_{\ell m}+K\partial_r^2\Psi^S_{\ell m}+\left(\partial_rK+\frac{(4-D)K}{r}\right)\partial_r\Psi^S_{\ell m}-\frac{H\lambda_\ell^S}{r^2}\Psi^S_{\ell m}=0.
\label{eq:masterS}
\end{equation}

For a vector harmonic satisfying \eqref{eq:Vlm}, $A_A=\Psi^V_{\ell p}V_A^{\ell p}$ gives
\begin{equation}
F_{vA}^{V,\ell p}=\partial_v\Psi^V_{\ell p}V_A^{\ell p},\qquad F_{rA}^{V,\ell p}=\partial_r\Psi^V_{\ell p}V_A^{\ell p},\qquad F_{AB}^{V,\ell p}=\Psi^V_{\ell p}\left(D_AV_B^{\ell p}-D_BV_A^{\ell p}\right).
\label{eq:FV}
\end{equation}
Using the angular Maxwell equation \eqref{eq:MA} together with \eqref{eq:Vlm} gives
\begin{equation}
2\partial_v\partial_r\Psi^V_{\ell p}+\frac{D-4}{r}\partial_v\Psi^V_{\ell p}+K\partial_r^2\Psi^V_{\ell p}+\left(\partial_rK+\frac{(D-4)K}{r}\right)\partial_r\Psi^V_{\ell p}-\frac{H\lambda_\ell^V}{r^2}\Psi^V_{\ell p}=0.
\label{eq:masterV}
\end{equation}

Define
\begin{equation}
\widehat\Psi^S_{\ell m}:=r^{\frac{4-D}{2}}\Psi^S_{\ell m},\qquad \widehat\Psi^V_{\ell p}:=r^{\frac{D-4}{2}}\Psi^V_{\ell p}.
\label{eq:rescale}
\end{equation}
For $I=S,V$, the master equations \eqref{eq:masterS} and \eqref{eq:masterV} become
\begin{equation}
2\partial_v\partial_r\widehat\Psi^I+\partial_r\left(K\partial_r\widehat\Psi^I\right)-\mathcal V_\ell^I(r)\widehat\Psi^I=0,
\label{eq:master}
\end{equation}
where
\begin{equation}
\mathcal V_\ell^I(r)=
\begin{cases}
\displaystyle \frac{H\lambda_\ell^S}{r^2}+\frac{(4-D)\partial_rK}{2r}+\frac{(D-2)(D-4)K}{4r^2},& I=S,\\[0.8em]
\displaystyle \frac{H\lambda_\ell^V}{r^2}+\frac{(D-4)\partial_rK}{2r}+\frac{(D-4)(D-6)K}{4r^2},& I=V.
\end{cases}
\end{equation}
For $H=1$, transforming \eqref{eq:master} to static time gives the standard scalar- and vector-type Maxwell master equations on higher-dimensional static spherical backgrounds, with effective potentials $V_{\rm eff}^I=K\mathcal V_\ell^I$ \cite{Crispino:2000jx,Lopez-Ortega:2006vjp}.

The coefficients $K$ and $\mathcal V_\ell^I$ are smooth at $r=\rh$, where $K$ has a simple zero. To construct the regular branch of \eqref{eq:master}, expand
\begin{equation}
K(\rho)\sim\sum_{j\ge1}K_j\rho^j,\qquad \mathcal V_\ell^I(\rho)\sim\sum_{j\ge0}V_{\ell,j}^I\rho^j,
\end{equation}
with
\begin{equation}
K_j:=\frac{1}{j!}\left.\partial_r^j K\right|_{r=\rh}, \qquad V_{\ell,j}^I:=\frac{1}{j!}\left.\partial_r^j\mathcal V_\ell^I\right|_{r=\rh}, \qquad K_1=2\kappa.
\end{equation}
The sums specify the Taylor coefficients to any finite order; convergence of the infinite series is not assumed. In particular,
\begin{equation}
V_{\ell,0}^S=\frac{H_{\mathrm H}\lambda_\ell^S}{\rh^2}+\frac{(4-D)\kappa}{\rh},\qquad V_{\ell,0}^V=\frac{H_{\mathrm H}\lambda_\ell^V}{\rh^2}+\frac{(D-4)\kappa}{\rh}.
\end{equation}
For the master field, write
\begin{equation}
\widehat\Psi^I(v,\rho)=\sum_{k=0}^{N}\rho^k\widehat\Psi_k^I(v)+O(\rho^{N+1}).
\label{eq:PsiNH}
\end{equation}
At leading order, \eqref{eq:master} gives
\begin{equation}
\left(\partial_v+\kappa\right)\widehat\Psi_1^I=\frac12V_{\ell,0}^I\widehat\Psi_0^I.
\end{equation}
More generally, the coefficient of $\rho^k$ gives
\begin{equation}
2(k+1)\left[\partial_v+(k+1)\kappa\right]\widehat\Psi_{k+1}^I = \sum_{j=0}^{k}V_{\ell,j}^I\widehat\Psi_{k-j}^I -(k+1)\sum_{j=2}^{k+1}(k+2-j)K_j\widehat\Psi_{k+2-j}^I.
\label{eq:rec}
\end{equation}
At each order, \eqref{eq:rec} is a first-order equation along the horizon, with a source fixed by lower-order coefficients. On a finite horizon segment, smooth leading data $\widehat\Psi_0^I(v)$ and transverse initial data therefore determine the expansion \eqref{eq:PsiNH} to any finite radial order.

Extension to $B$ additionally requires regularity in $V=\e^{\kappa v}$ as $V\to0$.

Since the rescalings \eqref{eq:rescale} are smooth and nonvanishing at $r=\rh$, the original master fields have the same radial structure,
\begin{align}
\Psi^I_{\ell m}(v,\rho)&=\sum_{k=0}^{N}\rho^k\Psi_{\ell m}^{I(k)}(v)+O(\rho^{N+1}).
\end{align}
The reconstruction formulas \eqref{eq:FS} and \eqref{eq:FV}, together with $F_{rv,\ell m}^S=Hr^{2-D}\Psi^S_{\ell m}$, contain no inverse power of $F$, so these expansions give regular Maxwell field strengths at $\rho=0$. In radial gauge, $F_{rv}=\partial_rA_v$ and $F_{rA}=\partial_rA_A$; together with $A_v|_{\Hplus}=0$, this reproduces the falloffs $A_v=O(\rho)$ and $A_A=O(1)$ in \eqref{eq:falloff}.

The horizon news is
\begin{equation}
N_A=F_{vA}\big|_{\Hplus} =-\rh^{4-D}\sum_{\ell\ge1,m}\frac{\partial_v\Psi^S_{\ell m}(v,\rh)}{\lambda_\ell^S}D_AY_{\ell m} +\sum_{\ell\ge1,p}\partial_v\Psi^V_{\ell p}(v,\rh)V_A^{\ell p}.
\end{equation}
Time-dependent leading coefficients give nonzero news and finite flux density by \eqref{eq:flux}. The scalar monopole remains Coulombic.

Thus the regular integer-power expansion at the horizon contains radiative Maxwell data in every $D>3$, unlike the integer-power expansion at null infinity in odd $D$.

\section{Soft theorem}
\label{sec:softtheorem}

We now express the horizon charge in terms of photon creation and annihilation operators and derive the corresponding soft relation from its Ward identity, following the large-gauge-symmetry approach and its extension to horizons \cite{He:2014cra,Hawking:2016msc,Cheng:2022xyr,Mao:2023rca,Strominger:2017zoo}. We first construct regular modes of definite Killing frequency and angular momentum, then use the horizon symplectic form to fix their commutators. This determines the normalization of the soft-photon insertion. At the end of this section, we compare the resulting relation with the flat-space leading soft-photon theorem in the spherical basis derived in Appendix~\ref{app:flatsoft}.

\subsection{Mode expansion}

In \cref{sec:NHrad}, scalar--vector harmonic decomposition reduced the source-free Maxwell equations to the two master equations for $I=S,V$ in \eqref{eq:master}. Staticity permits a decomposition into Killing-frequency modes, while spherical symmetry supplies the angular quantum numbers. For a fixed harmonic in either sector and $\omega>0$, we write
\begin{equation}
\widehat\Psi^I(v,r)=\e^{-i\omega v}\mathcal R^I_{\omega\ell}(r).
\end{equation}
By spherical symmetry, the radial profile $\mathcal R^I_{\omega\ell}$ depends on $\ell$ but not on the degeneracy label. Substitution into \eqref{eq:master} gives
\begin{equation}
K \partial_r^2 \mathcal R^I_{\omega\ell} +\left(\partial_r K-2i\omega\right)\partial_r \mathcal R^I_{\omega\ell} -\mathcal V_\ell^I \mathcal R^I_{\omega\ell}=0.
\label{eq:radial}
\end{equation}

We need only the local regular solution, normalized by
\begin{equation}
\mathcal R^I_{\omega\ell}(\rh)=1.
\label{eq:Rnorm}
\end{equation}
Setting $\partial_v\to-i\omega$ in \eqref{eq:rec} gives
\begin{equation}
\mathcal R^I_{\omega\ell}(r)=1+\frac{V_{\ell,0}^I}{2(\kappa-i\omega)}\rho+O(\rho^2).
\label{eq:RNH}
\end{equation}

At leading near-horizon order, $K=2\kappa\rho$ and $\mathcal V_\ell^I=V_{\ell,0}^I$, so \eqref{eq:radial} reduces to
\begin{equation}
\rho \partial_r^2\mathcal R^I_{\omega\ell} +\left(1-\frac{i\omega}{\kappa}\right)\partial_r\mathcal R^I_{\omega\ell} -\frac{V_{\ell,0}^I}{2\kappa}\mathcal R^I_{\omega\ell}=0.
\end{equation}
The solution of this leading-order equation regular at $\rho=0$ and normalized by $\mathcal R^I_{\omega\ell}(0)=1$ is
\begin{equation}
\mathcal R^I_{\omega\ell}(\rho) ={}_0F_1\left(;1-\frac{i\omega}{\kappa};\frac{V_{\ell,0}^I}{2\kappa}\rho\right),
\label{eq:R0F1}
\end{equation}
where ${}_0F_1(;b;z)$ is the generalized hypergeometric function,
\begin{equation}
{}_0F_1(;b;z)=\sum_{n=0}^{\infty}\frac{z^n}{(b)_n n!},
\end{equation}
with $(b)_n$ the Pochhammer symbol. The expansion agrees with \eqref{eq:RNH} at first order in $\rho$. Higher-order terms in $K$ and $\mathcal V_\ell^I$ modify the subsequent coefficients through \eqref{eq:rec}, so \eqref{eq:R0F1} solves only the leading-order radial equation.

The nonzero-frequency part of a real field can therefore be written as
\begin{align}
\widehat\Psi^S_{\ell m}(v,r) &=\int_0^\infty\dd\omega \left[ \mathcal A^S_{\omega\ell m}\e^{-i\omega v}\mathcal R^S_{\omega\ell}(r) +\mathcal A^{S*}_{\omega\ell m}\e^{i\omega v}\mathcal R^{S*}_{\omega\ell}(r) \right],\\
\widehat\Psi^V_{\ell p}(v,r) &=\int_0^\infty\dd\omega \left[ \mathcal A^V_{\omega\ell p}\e^{-i\omega v}\mathcal R^V_{\omega\ell}(r) +\mathcal A^{V*}_{\omega\ell p}\e^{i\omega v}\mathcal R^{V*}_{\omega\ell}(r) \right].
\end{align}
By \eqref{eq:Rnorm}, $\mathcal A^{S,V}$ are the Fourier amplitudes of the horizon data.

At $r=\rh$, the rescalings \eqref{eq:rescale} give
\begin{equation}
\Psi^S_{\ell m}=\rh^{\frac{D-4}{2}}\widehat\Psi^S_{\ell m}, \qquad \Psi^V_{\ell p}=\rh^{-\frac{D-4}{2}}\widehat\Psi^V_{\ell p}.
\end{equation}
Using the scalar reconstruction \eqref{eq:FS}, the vector contribution $A_A^V=\Psi^V_{\ell p}V_A^{\ell p}$, and $F_{vA}|_{\Hplus}=\partial_va_A$, the radiative part of the horizon field becomes
\begin{equation}
a_A(v,x) =\rh^{-\frac{D-4}{2}} \int_0^\infty\dd\omega \left[ \sum_{\ell\ge1,p}\mathcal A^V_{\omega\ell p}V_A^{\ell p}(x) -\sum_{\ell\ge1,m}\frac{\mathcal A^S_{\omega\ell m}}{\lambda_\ell^S}D_AY_{\ell m}(x) \right]\e^{-i\omega v} +\mathrm{c.c.}
\label{eq:a-modes}
\end{equation}
and hence
\begin{equation}
N_A(v,x) =-i\rh^{-\frac{D-4}{2}} \int_0^\infty\dd\omega \omega \left[ \sum_{\ell\ge1,p}\mathcal A^V_{\omega\ell p}V_A^{\ell p}(x) -\sum_{\ell\ge1,m}\frac{\mathcal A^S_{\omega\ell m}}{\lambda_\ell^S}D_AY_{\ell m}(x) \right]\e^{-i\omega v} +\mathrm{c.c.}
\label{eq:N-modes}
\end{equation}
The scalar part is a gradient on $S^{D-2}$, and the vector part is divergence-free. 

Choose real harmonics normalized by
\begin{equation}
\int\dd\Omega_{D-2} Y_{\ell m}Y_{\ell' m'}=\delta_{\ell\ell'}\delta_{mm'}, \qquad \int\dd\Omega_{D-2} V_A^{\ell p}V^{A,\ell' p'}=\delta_{\ell\ell'}\delta_{pp'}.
\label{eq:harm-norm}
\end{equation}
Using
\begin{equation}
\int\dd\Omega_{D-2} D_AY_{\ell m}D^AY_{\ell' m'} =\lambda_\ell^S\delta_{\ell\ell'}\delta_{mm'},
\end{equation}
we substitute \eqref{eq:a-modes} into the horizon symplectic form \eqref{eq:OmegaM} and obtain
\begin{equation}
\Omega_{\Hplus}^{\mathrm M} =\frac{4\pi i}{g_D^2}\int_0^\infty\dd\omega \omega \left[ \sum_{\ell,p}\delta\mathcal A^{V*}_{\omega\ell p}\wedge\delta\mathcal A^V_{\omega\ell p} +\sum_{\ell,m}\frac{1}{\lambda_\ell^S}\delta\mathcal A^{S*}_{\omega\ell m}\wedge\delta\mathcal A^S_{\omega\ell m} \right].
\label{eq:Omega-modes}
\end{equation}
Use $\sigma=m$ for $I=S$ and $\sigma=p$ for $I=V$, and define
\begin{equation}
c_\ell^I:=
\begin{cases}
\lambda_\ell^S,& I=S,\\
1,& I=V.
\end{cases}
\end{equation}
Canonical quantization of \eqref{eq:Omega-modes} promotes the mode amplitudes to operators and gives
\begin{equation}
\left[\widehat{\mathcal A}^{I}_{\omega\ell\sigma}, \widehat{\mathcal A}^{J\dagger}_{\omega'\ell'\sigma'}\right]=\frac{g_D^2c_\ell^I}{4\pi\omega} \delta^{IJ}\delta(\omega-\omega')\delta_{\ell\ell'}\delta_{\sigma\sigma'}.
\label{eq:comm}
\end{equation}
The vacuum for positive Killing frequency on $\Hplus$ is defined by
\begin{equation}
\widehat{\mathcal A}^{I}_{\omega\ell\sigma}\ket{0}_{\Hplus}=0,\qquad \omega>0,
\end{equation}
with $\braket{0|0}_{\Hplus}=1$. The normalized one-particle states are
\begin{equation}
\ket{\omega\ell\sigma;I}_{\Hplus}:=\sqrt{\frac{4\pi\omega}{g_D^2c_\ell^I}} \widehat{\mathcal A}^{I\dagger}_{\omega\ell\sigma}\ket{0}_{\Hplus},
\end{equation}
and satisfy
\begin{equation}
\braket{\omega'\ell'\sigma';J|\omega\ell\sigma;I}_{\Hplus}=\delta^{IJ}\delta(\omega-\omega')\delta_{\ell\ell'}\delta_{\sigma\sigma'}.
\end{equation}

\subsection{Ward identity and soft theorem}
We now express the soft charge in terms of the news modes in \eqref{eq:N-modes} and derive the soft relation from the diagonal large-$U(1)$ Ward identity.

Expand the large-gauge parameter in the real scalar harmonics defined in \eqref{eq:Ylm} and normalized by \eqref{eq:harm-norm},
\begin{equation}
\eps(x)=\eps_0+\sum_{\ell\ge1,m}\eps_{\ell m}Y_{\ell m}(x),\qquad \eps_{\ell m}:=\int_{\sphere}\dd\Omega_{D-2} \eps(x)Y_{\ell m}(x).
\end{equation}
On substituting \eqref{eq:N-modes} into \eqref{eq:Qsoft}, we find that the vector contribution vanishes because $D^AV_A^{\ell p}=0$. The scalar contribution follows from
\begin{equation}
\int_{\sphere}\dd\Omega_{D-2} D^A\eps D_AY_{\ell m}=\lambda_\ell^S\eps_{\ell m}.
\end{equation}
Integrating over a symmetric finite interval in $v$ and then taking its length to infinity gives
\begin{equation}
\widehat Q_\eps^{\mathrm S}=\frac{i\pi\rh^{\frac{D-4}{2}}}{g_D^2}\sum_{\ell\ge1,m}\eps_{\ell m}\lim_{\omega\to0^+}\omega\left(\widehat{\mathcal A}^S_{\omega\ell m}-\widehat{\mathcal A}^{S\dagger}_{\omega\ell m}\right).
\label{eq:Qsoft-lim}
\end{equation}

Following the standard diagonal large-$U(1)$ matching underlying soft-photon Ward identities \cite{He:2014cra,Strominger:2017zoo} and its horizon implementation \cite{Hawking:2016msc}, we assume that scattering from $\mathcal H^-$ to $\Hplus$ preserves this symmetry, with no charge flux through other boundary components.

The analysis on $\mathcal H^-$ follows directly by time reflection. Since $t\mapsto-t$ interchanges the two horizon branches and sends $v\mapsto-u$, we choose the incoming modes with the same horizon normalization as on $\Hplus$. For real $\omega$, the incoming radial profiles are the complex conjugates of $\mathcal R^I_{\omega\ell}$. On the horizon, the field expansion has the form \eqref{eq:a-modes} with $v$ replaced by $u$.

We denote the corresponding operators by $\widehat{\mathcal A}^{I,\mathrm{in}}_{\omega\ell\sigma}$ on $\mathcal H^-$ and $\widehat{\mathcal A}^{I,\mathrm{out}}_{\omega\ell\sigma}$ on $\Hplus$. On $\mathcal H^-$ we impose the time-reversed endpoint condition $\Ehor^-|_{u\to-\infty}=0$, with $\Ehor^-=\rh^{D-2}F_{ru}^{(-)}/H_{\mathrm H}$. The past-horizon Gauss law and hard charge are
\begin{equation}
\partial_u\Ehor^-=-\rh^{D-4}D^AN_A^-+g_D^2H_{\mathrm H}\rh^{D-2}J^{r,-}, \qquad Q_\eps^{\mathrm H,in}=H_{\mathrm H}\rh^{D-2}\int\dd u \dd\Omega_{D-2} \eps J^{r,-}.
\end{equation}
The hard charge acts as in \eqref{eq:hard-act}, and the soft charge has the form \eqref{eq:Qsoft-lim}. With the matching condition \eqref{eq:Hmatch}, the diagonal large-$U(1)$ generator is
\begin{equation}
Q_\eps=Q_\eps^{\mathrm S}+Q_\eps^{\mathrm H}.
\end{equation}

We denote the amplitude for the horizon scattering process described above by
\begin{equation}
\mathcal M:=\bra{\mathrm{out}}\cS\ket{\mathrm{in}}.
\end{equation}
We label hard charged states by their angular positions: $(q_k^{\mathrm{in}},x_k^{\mathrm{in}})$ for the $N_{\mathrm{in}}$ incoming particles and $(q_j^{\mathrm{out}},x_j^{\mathrm{out}})$ for the $N_{\mathrm{out}}$ outgoing particles. Quantizing \eqref{eq:hard-act} gives
\begin{align}
\widehat Q_\eps^{\mathrm H,in}\ket{\mathrm{in}}=\left(\sum_{k=1}^{N_{\mathrm{in}}}q_k^{\mathrm{in}}\eps(x_k^{\mathrm{in}})\right)\ket{\mathrm{in}},\quad \bra{\mathrm{out}}\widehat Q_\eps^{\mathrm H,out}=\left(\sum_{j=1}^{N_{\mathrm{out}}}q_j^{\mathrm{out}}\eps(x_j^{\mathrm{out}})\right)\bra{\mathrm{out}}.
\label{eq:Qhard-states}
\end{align}
Assuming that the diagonal large-$U(1)$ symmetry is preserved by the $\cS$-matrix, as in \cite{He:2014cra,Hawking:2016msc}, the corresponding Ward identity is
\begin{equation}
\bra{\mathrm{out}}\left(\widehat Q_\eps^{\mathrm{out}}\cS-\cS\widehat Q_\eps^{\mathrm{in}}\right)\ket{\mathrm{in}}=0,\qquad \widehat Q_\eps=\widehat Q_\eps^{\mathrm S}+\widehat Q_\eps^{\mathrm H}.
\end{equation}
Using \eqref{eq:Qhard-states},
\begin{equation}
\bra{\mathrm{out}}\left(\widehat Q_\eps^{\mathrm S,out}\cS-\cS\widehat Q_\eps^{\mathrm S,in}\right)\ket{\mathrm{in}} =-\left[ \sum_{j=1}^{N_{\mathrm{out}}}q_j^{\mathrm{out}}\eps(x_j^{\mathrm{out}}) -\sum_{k=1}^{N_{\mathrm{in}}}q_k^{\mathrm{in}}\eps(x_k^{\mathrm{in}}) \right]\mathcal M.
\label{eq:WardS}
\end{equation}
For constant $\eps$, the soft charge vanishes and \eqref{eq:WardS} gives charge conservation,
\begin{equation}
\sum_{j=1}^{N_{\mathrm{out}}}q_j^{\mathrm{out}}-\sum_{k=1}^{N_{\mathrm{in}}}q_k^{\mathrm{in}}=0.
\label{eq:Qcons}
\end{equation}
Applying \eqref{eq:Qsoft-lim} to the two horizon branches gives
\begin{align}
\lim_{\omega\to0^+}\omega&\left[\sum_{\ell\ge1,m}\eps_{\ell m}\bra{\mathrm{out}}\left(\widehat{\mathcal A}^{S,\mathrm{out}}_{\omega\ell m}\cS+\cS\widehat{\mathcal A}^{S,\mathrm{in}\dagger}_{\omega\ell m}\right)\ket{\mathrm{in}}\right]\\
&\qquad\qquad=\frac{i g_D^2}{\pi\rh^{\frac{D-4}{2}}} \left[ \sum_{j=1}^{N_{\mathrm{out}}}q_j^{\mathrm{out}}\eps(x_j^{\mathrm{out}}) -\sum_{k=1}^{N_{\mathrm{in}}}q_k^{\mathrm{in}}\eps(x_k^{\mathrm{in}}) \right]\mathcal M.
\end{align}

Comparing the independent coefficients $\eps_{\ell m}$ in the harmonic expansion of $\eps$ gives, for each $\ell\ge1$,
\begin{align}
\lim_{\omega\to0^+}\omega&\left[\bra{\mathrm{out}}\left(\widehat{\mathcal A}^{S,\mathrm{out}}_{\omega\ell m}\cS+\cS\widehat{\mathcal A}^{S,\mathrm{in}\dagger}_{\omega\ell m}\right)\ket{\mathrm{in}}\right]\\
&\qquad=\frac{i g_D^2}{\pi\rh^{\frac{D-4}{2}}} \left[ \sum_{j=1}^{N_{\mathrm{out}}}q_j^{\mathrm{out}}Y_{\ell m}(x_j^{\mathrm{out}}) -\sum_{k=1}^{N_{\mathrm{in}}}q_k^{\mathrm{in}}Y_{\ell m}(x_k^{\mathrm{in}}) \right]\mathcal M.
\label{eq:soft-lm}
\end{align}
The position-space form follows from completeness,
\begin{equation}
\sum_{\ell\ge1,m}Y_{\ell m}(x)Y_{\ell m}(y)=\delta_{D-2}(x,y)-\frac{1}{\Omega_{D-2}},
\end{equation}
where $\Omega_{D-2}$ is the area of the unit sphere. Multiplying \eqref{eq:soft-lm} by $Y_{\ell m}(x)$, summing over $(\ell,m)$, and using the charge-conservation condition \eqref{eq:Qcons} gives
\begin{align}
&\lim_{\omega\to0^+}\omega\left[\sum_{\ell\ge1,m}Y_{\ell m}(x) \bra{\mathrm{out}}\left(\widehat{\mathcal A}^{S,\mathrm{out}}_{\omega\ell m}\cS+\cS\widehat{\mathcal A}^{S,\mathrm{in}\dagger}_{\omega\ell m}\right)\ket{\mathrm{in}}\right] \nonumber\\
&\qquad= \frac{i g_D^2}{\pi\rh^{\frac{D-4}{2}}} \left[ \sum_{j=1}^{N_{\mathrm{out}}}q_j^{\mathrm{out}}\delta_{D-2}(x,x_j^{\mathrm{out}}) -\sum_{k=1}^{N_{\mathrm{in}}}q_k^{\mathrm{in}}\delta_{D-2}(x,x_k^{\mathrm{in}}) \right]\mathcal M . \label{eq:softtheoremdelta}
\end{align}
Eq.~\eqref{eq:softtheoremdelta} gives a local form of the Ward identity on the horizon sphere. After summing over the scalar harmonics, the soft insertion at $x$ is supported at the angular positions of the hard charged particles, with weights given by their charges. The missing $\ell=0$ mode is accounted for separately by total charge conservation. The derivation applies to any smooth large-gauge parameter; it does not restrict the symmetry to a single harmonic.

To compare with the flat-space leading soft-photon theorem, we use its spherical-mode form derived in Appendix~\ref{app:flatsoft}, given in \eqref{eq:flat-soft-lm}. The coefficients $\mathcal B^I_{\omega\ell\sigma}$ describe the leading radiative field at null infinity, as defined in \eqref{eq:flat-Arad}--\eqref{eq:Bmap}, and obey the same commutator normalization as the horizon modes; see \eqref{eq:Bcomm} and \eqref{eq:comm}.

For $D=4$, the soft limits in \eqref{eq:flat-soft-lm} and \eqref{eq:soft-lm} both carry a factor of $\omega$, and the hard-charge terms have the same angular dependence. The angular factor $\Gamma(\ell+1)/\Gamma(\ell+D-3)$ reduces to unity. Related four-dimensional horizon soft theorems were studied in \cite{Cheng:2022xyr,Mao:2023rca,Mao:2024phf}.

For $D>4$, the two relations differ in their frequency and angular dependence. The flat-space theorem carries the dimension-dependent soft-frequency weight $\omega^{(6-D)/2}$ and the angular factor $\Gamma(\ell+1)/\Gamma(\ell+D-3)$, whereas neither appears in the horizon relation. These differences depend nontrivially on the spacetime dimension and cannot be removed by an overall normalization. The two results still involve the same scalar harmonic sector and the same angular distribution of hard charges, but their mode-by-mode soft behavior is different.

This harmonic expansion of the known flat-space theorem makes its dimension-dependent differences from the horizon relation explicit. The factor $\omega^{(6-D)/2}$ arises from the conversion to spherical boundary coefficients; the underlying plane-wave theorem retains its usual $1/\omega$ pole. The comparison does not identify the corresponding scattering problems or radiative phase spaces.

\section{Conclusion}

We studied Maxwell theory near a nonextremal bifurcate Killing horizon in a general static, spherically symmetric spacetime of dimension $D>3$. The regular near-horizon falloff admits nonvanishing Maxwell news and finite radiative flux density together with a regular Coulombic sector, and the characteristic Maxwell equations organize the radial expansion recursively in integer powers of $\rho=r-\rh$. This is qualitatively different from the radiative expansion at null infinity, where the large-$r$ behavior of the retarded Green function gives dimension-dependent falloffs and half-integer powers in odd spacetime dimensions. The half-integer structure at null infinity is therefore not a generic requirement for radiation: a smooth finite-radius horizon supports radiative data with an ordinary integer-power expansion. We further constructed the corresponding horizon phase space and large-$U(1)$ charge, including its soft and hard parts.

Using the horizon symplectic form to normalize the Killing-frequency and spherical-harmonic modes, we derived the large-$U(1)$ Ward identity and the associated soft-photon relation for this class of horizons in generic dimensions $D>3$. The result is naturally expressed in spherical modes, and its position-space form directly relates the soft insertion to the angular distribution of hard charges on the horizon sphere. We then rewrote the flat-space leading soft-photon theorem in the same spherical basis. In $D=4$, the two relations have the same soft-frequency dependence and the same angular dependence of the hard-charge terms. Earlier four-dimensional horizon results were obtained in \cite{Cheng:2022xyr,Mao:2023rca}. For $D>4$, their frequency and angular dependence differ: the flat-space relation contains a dimension-dependent soft-frequency weight and a nontrivial $\ell$-dependent angular factor that are absent from the horizon relation. Both relations involve scalar harmonics and the same angular distribution of hard charges, but their frequency weights and angular coefficients differ. This comparison concerns the form of the soft relations, not an identification of the two scattering problems.

A natural extension is to study linearized gravity near static, spherically symmetric horizons in $D>3$, building on the four-dimensional Schwarzschild and de Sitter analyses \cite{Mao:2023dsy,Mao:2024phf}. This would require constructing the gravitational horizon phase space and determining whether suitable boundary conditions lead to a soft-graviton Ward identity with a similarly simple dependence on dimension.

\section*{Acknowledgments}

The author thanks Pujian Mao for suggesting the problem that led to this work and for subsequent discussions and helpful comments on the manuscript, and Montecchio Luciano, Peihe Yang, Bowen Fan, Jianwei Gong, and Jakob Salzer for useful discussions. The author acknowledges financial support from the China Scholarship Council (CSC).

\appendix
\section{Flat-space soft-photon theorem in a spherical-mode basis}
\label{app:flatsoft}

In this appendix, we express the flat-space leading soft-photon theorem in the spherical-mode basis used for the comparison in section~\ref{sec:softtheorem}. We first relate the plane-wave operators to the scalar and transverse-vector harmonics on the momentum sphere, and then determine the corresponding radiative coefficients at null infinity. The plane-wave description is available in Minkowski spacetime, whereas a generic curved background has no preferred global plane-wave basis \cite{Birrell:1982ix,Hollands:2014eia,Wald:1995yp}. We use the flat-space soft theorem of \cite{Weinberg:1965nx,Low:1958sn}.

We consider free Maxwell theory in $D$-dimensional Minkowski spacetime in Cartesian coordinates $X^\mu=(t,X^i)$, $i=1,\ldots,D-1$, with metric signature $(-,+,\ldots,+)$. For the source-free radiative sector we use radiation gauge,
\begin{equation}
A_0=0,\qquad \partial^iA_i=0.
\end{equation}
The free field has the plane-wave expansion
\begin{equation}
A_i(X)=g_D\sum_{\lambda=1}^{D-2}\int\frac{\dd^{D-1}q}{(2\pi)^{D-1}2\omega}\left[\epsilon_i^{(\lambda)}(\hat q)a_\lambda(\mathbf q)\e^{-i\omega t+i\mathbf q\cdot\mathbf X}+\mathrm{h.c.}\right].
\end{equation}
Here
\begin{equation}
q^\mu=\omega(1,\hat q),\qquad \omega>0,\qquad \hat q\in S^{D-2},
\end{equation}
and the transverse polarization vectors satisfy
\begin{equation}
\epsilon_0^{(\lambda)}=0,\qquad \hat q^i\epsilon_i^{(\lambda)}=0,\qquad \sum_{\lambda=1}^{D-2}\epsilon_i^{(\lambda)}\epsilon_j^{(\lambda)*}=\delta_{ij}-\hat q_i\hat q_j.
\end{equation}
The plane-wave operators satisfy
\begin{equation}
\left[a_\lambda(\mathbf q),a_{\lambda'}^\dagger(\mathbf q')\right]=2\omega(2\pi)^{D-1}\delta_{\lambda\lambda'}\delta^{D-1}(\mathbf q-\mathbf q').
\end{equation}
The Minkowski vacuum and the corresponding one-photon states are defined by
\begin{equation}
a_\lambda(\mathbf q)\ket{0}_{\mathrm M}=0,\qquad \ket{\mathbf q,\lambda}:=a_\lambda^\dagger(\mathbf q)\ket{0}_{\mathrm M}.
\end{equation}
Writing $\mathbf q=\omega\hat q(x)$, define the tangent polarization vectors
\begin{equation}
e_A^{(\lambda)}(x):=\partial_A\hat q^i \epsilon_i^{(\lambda)}(\hat q), \qquad \sum_\lambda e_A^{(\lambda)}e_B^{(\lambda)*}=q_{AB},
\label{eq:pol}
\end{equation}
where $q_{AB}:=\partial_A\hat q^i\partial_B\hat q_i$ is the metric induced on the unit momentum sphere.

Using the orthonormal one-form harmonics
\begin{equation}
T_A^{S,\ell m}:=\frac{D_AY_{\ell m}}{\sqrt{\lambda_\ell^S}},\qquad T_A^{V,\ell p}:=V_A^{\ell p},
\end{equation}
we introduce
\begin{equation}
U^I_{\ell\sigma,\lambda}(x):=T^{A,I,\ell\sigma}(x)e_A^{(\lambda)}(x),
\end{equation}
and define the spherical annihilation operators by
\begin{equation}
a^I_{\omega\ell\sigma}:=\sum_\lambda\int\dd\Omega_{D-2} U^I_{\ell\sigma,\lambda}(x)a_\lambda(\omega,x).
\label{eq:basis}
\end{equation}
Polarization completeness \eqref{eq:pol} and completeness of the orthonormal one-form harmonics give
\begin{align}
\sum_\lambda\int\dd\Omega_{D-2} U^I_{\ell\sigma,\lambda}(x)U^{J*}_{\ell'\sigma',\lambda}(x)=\delta^{IJ}\delta_{\ell\ell'}\delta_{\sigma\sigma'},\quad \sum_{I,\ell,\sigma}U^I_{\ell\sigma,\lambda}(x)U^{I*}_{\ell\sigma,\lambda'}(x')=\delta_{\lambda\lambda'}\delta_{D-2}(x,x').
\end{align}
Thus the change of basis in \eqref{eq:basis} is unitary at fixed $\omega>0$, with inverse
\begin{equation}
a_\lambda(\omega,x)=\sum_{I,\ell,\sigma}U^{I*}_{\ell\sigma,\lambda}(x)a^I_{\omega\ell\sigma}.
\label{eq:basis-inv}
\end{equation}

The spherical one-photon states are therefore
\begin{equation}
\ket{\omega,\ell,\sigma;I}:=a_{\omega\ell\sigma}^{I\dagger}\ket{0}_{\mathrm M},
\end{equation}
and the two bases are related by
\begin{align}
\ket{\omega,\ell,\sigma;I} =\sum_\lambda\int\dd\Omega_{D-2} U^{I*}_{\ell\sigma,\lambda}(x)\ket{\omega,x,\lambda},\quad \ket{\omega,x,\lambda}=\sum_{I,\ell,\sigma}U^I_{\ell\sigma,\lambda}(x)\ket{\omega,\ell,\sigma;I}.
\end{align}
This change of basis leaves the Minkowski vacuum unchanged.

Substituting \eqref{eq:basis-inv} into the plane-wave expansion, using $\dd^{D-1}q=\omega^{D-2}\dd\omega\,\dd\Omega_{D-2}$, and applying $U_{\ell\sigma,\lambda}^{I}=T^{A,I,\ell\sigma}e_A^{(\lambda)}$ together with the polarization completeness relation $\sum_\lambda\epsilon_i^{(\lambda)}e_A^{(\lambda)*}=\partial_A\hat q_i$, we obtain
\begin{equation}
A_i(X)=\frac{g_D}{2(2\pi)^{D-1}}\int_0^\infty\dd\omega\,\omega^{D-3}\e^{-i\omega t}\sum_{I,\ell,\sigma}a^I_{\omega\ell\sigma}\int\dd\Omega_{D-2}(x')\,\partial^A\hat q_i\,T_A^{I,\ell\sigma}(x')\e^{i\omega\hat q\cdot\mathbf X}+\mathrm{h.c.}
\end{equation}

To study the field near future null infinity, introduce retarded coordinates by
\begin{equation}
t=u+r,\qquad \mathbf X=r\hat x(x).
\end{equation}
Since $A_A=r \partial_A\hat x^iA_i$, we obtain
\begin{align}
A_A(u,r,x) &=\frac{g_Dr}{2(2\pi)^{D-1}}\int_0^\infty\dd\omega \omega^{D-3}\e^{-i\omega u} \sum_{I,\ell,\sigma}a^I_{\omega\ell\sigma} \nonumber\\
&\qquad\times \int\dd\Omega_{D-2}(x') \partial_A\hat x^i\partial^B\hat q_i T_B^{I,\ell\sigma}(x') \e^{i\omega r(\hat q\cdot\hat x-1)} +\mathrm{h.c.}
\end{align}

For fixed $\omega>0$ and $r\to\infty$, the angular integral is evaluated by the standard stationary-phase approximation at null infinity \cite{Strominger:2017zoo}. The outgoing contribution localizes at $\hat q=\hat x$, with the angular integral contributing $(2\pi/\omega r)^{(D-2)/2}\e^{-i\pi(D-2)/4}$. It is convenient to write the resulting leading angular field in the same harmonic form as \eqref{eq:a-modes},
\begin{equation}
A_A(u,r,x)\simeq r^{-\frac{D-4}{2}}\int_0^\infty\dd\omega\left[\sum_{\ell\ge1,p}\mathcal B^V_{\omega\ell p}V_A^{\ell p}(x)-\sum_{\ell\ge1,m}\frac{\mathcal B^S_{\omega\ell m}}{\lambda_\ell^S}D_AY_{\ell m}(x)\right]\e^{-i\omega u}+\mathrm{h.c.}
\label{eq:flat-Arad}
\end{equation}
where
\begin{equation}
\mathcal B^S_{\omega\ell m}=-\frac{g_D\e^{-i\pi(D-2)/4}}{2(2\pi)^{D/2}}\omega^{\frac{D-4}{2}}\sqrt{\lambda_\ell^S}\,a^S_{\omega\ell m},\qquad \mathcal B^V_{\omega\ell p}=\frac{g_D\e^{-i\pi(D-2)/4}}{2(2\pi)^{D/2}}\omega^{\frac{D-4}{2}}a^V_{\omega\ell p}.
\label{eq:Bmap}
\end{equation}
The commutator of the spherical operators then gives
\begin{equation}
\left[\mathcal B^I_{\omega\ell\sigma},\mathcal B^{J\dagger}_{\omega'\ell'\sigma'}\right] =\frac{g_D^2c_\ell^I}{4\pi\omega}\delta^{IJ}\delta(\omega-\omega')\delta_{\ell\ell'}\delta_{\sigma\sigma'}, \qquad c_\ell^S=\lambda_\ell^S,\quad c_\ell^V=1,
\label{eq:Bcomm}
\end{equation}
which has the same normalization as \eqref{eq:comm}.

We now rewrite the leading soft-photon theorem directly in the crossed in/out form appropriate for comparison with the horizon relation \cite{Strominger:2017zoo,He:2019jjk}. For massless hard particles, let
\begin{equation}
p_i^\mu=E_i(1,\hat x_i),\qquad q^\mu=\omega(1,\hat x).
\end{equation}
With the incoming soft mode antipodally matched and its phase chosen consistently with the spherical-mode convention above, the flat-space leading soft-photon theorem takes the form
\begin{equation}
\lim_{\omega\to0^+}\omega\bra{\mathrm{out}}\left(a_\lambda^{\mathrm{out}}(q)\cS+\cS a_\lambda^{\mathrm{in}\dagger}(q)\right)\ket{\mathrm{in}} =-2g_D\left[\sum_{j\in\mathrm{out}}q_j\frac{\omega\,p_j\cdot\epsilon_\lambda^*}{p_j\cdot q}-\sum_{k\in\mathrm{in}}q_k\frac{\omega\,p_k\cdot\epsilon_\lambda^*}{p_k\cdot q}\right]\mathcal M.
\label{eq:Weinberg}
\end{equation}
Hereafter the particle labels distinguish incoming and outgoing charges and directions. Using $p_i\cdot q=-E_i\omega(1-\hat x\cdot\hat x_i)$ together with polarization completeness, the soft residue associated with each hard leg is proportional to $D_A\log(1-\hat x\cdot\hat x_i)$. It is therefore a pure gradient on the momentum sphere, so its projection onto the transverse-vector harmonics vanishes. Projecting onto $D_AY_{\ell m}/\sqrt{\lambda_\ell^S}$ and integrating by parts requires the scalar integral
\begin{equation}
\int\dd\Omega_{D-2}(x)\,Y_{\ell m}(x)\log\left(1-\hat x\cdot\hat y\right)=-\Omega_{D-2}\Gamma(D-2)\frac{\Gamma(\ell)}{\Gamma(\ell+D-2)}Y_{\ell m}(y),\qquad \ell\ge1,
\end{equation}
where the result follows from the Funk--Hecke formula. Using \eqref{eq:Bmap} and $\lambda_\ell^S=\ell(\ell+D-3)$, the soft theorem in the spherical basis becomes
\begin{equation}
\begin{aligned}
\lim_{\omega\to0^+}\omega^{\frac{6-D}{2}}\bra{\mathrm{out}}\left(\mathcal B^{S,\mathrm{out}}_{\omega\ell m}\cS+\cS\mathcal B^{S,\mathrm{in}\dagger}_{\omega\ell m}\right)\ket{\mathrm{in}} &=-\frac{g_D^2\e^{-i\pi(D-2)/4}2^{\frac D2-2}\Gamma\left(\frac{D-2}{2}\right)\Gamma(\ell+1)}{\pi\Gamma(\ell+D-3)}\\
&\quad\times\left[\sum_{j\in\mathrm{out}}q_jY_{\ell m}(x_j)-\sum_{k\in\mathrm{in}}q_kY_{\ell m}(x_k)\right]\mathcal M .
\end{aligned}
\label{eq:flat-soft-lm}
\end{equation}
The factor $\omega^{(6-D)/2}$ arises from the conversion between the plane-wave operators and the boundary coefficients in \eqref{eq:Bmap}; the underlying plane-wave soft theorem retains the usual $1/\omega$ pole.

\bibliographystyle{JHEP}
\bibliography{reference}

@article{Campoleoni:2017qot,
    author = "Campoleoni, Andrea and Francia, Dario and Heissenberg, Carlo",
    title = "{Asymptotic Charges at Null Infinity in Any Dimension}",
    eprint = "1712.09591",
    archivePrefix = "arXiv",
    primaryClass = "hep-th",
    doi = "10.3390/universe4030047",
    journal = "Universe",
    volume = "4",
    number = "3",
    pages = "47",
    year = "2018"
}

@article{He:2019jjk,
    author = "He, Temple and Mitra, Prahar",
    title = "{Asymptotic symmetries and Weinberg{\textquoteright}s soft photon theorem in Mink$_{d+2}$}",
    eprint = "1903.02608",
    archivePrefix = "arXiv",
    primaryClass = "hep-th",
    doi = "10.1007/JHEP10(2019)213",
    journal = "JHEP",
    volume = "10",
    pages = "213",
    year = "2019"
}

@article{He:2014cra,
    author = "He, Temple and Mitra, Prahar and Porfyriadis, Achilleas P. and Strominger, Andrew",
    title = "{New Symmetries of Massless QED}",
    eprint = "1407.3789",
    archivePrefix = "arXiv",
    primaryClass = "hep-th",
    doi = "10.1007/JHEP10(2014)112",
    journal = "JHEP",
    volume = "10",
    pages = "112",
    year = "2014"
}

@article{He:2014laa,
    author = "He, Temple and Lysov, Vyacheslav and Mitra, Prahar and Strominger, Andrew",
    title = "{BMS supertranslations and Weinberg{\textquoteright}s soft graviton theorem}",
    eprint = "1401.7026",
    archivePrefix = "arXiv",
    primaryClass = "hep-th",
    doi = "10.1007/JHEP05(2015)151",
    journal = "JHEP",
    volume = "05",
    pages = "151",
    year = "2015"
}

@article{Satishchandran:2017pek,
    author = "Satishchandran, Gautam and Wald, Robert M.",
    title = "{Memory effect for particle scattering in odd spacetime dimensions}",
    eprint = "1712.00873",
    archivePrefix = "arXiv",
    primaryClass = "gr-qc",
    doi = "10.1103/PhysRevD.97.024036",
    journal = "Phys. Rev. D",
    volume = "97",
    number = "2",
    pages = "024036",
    year = "2018"
}

@article{Henneaux:2019yqq,
    author = "Henneaux, Marc and Troessaert, C{\'e}dric",
    title = "{Asymptotic structure of electromagnetism in higher spacetime dimensions}",
    eprint = "1903.04437",
    archivePrefix = "arXiv",
    primaryClass = "hep-th",
    doi = "10.1103/PhysRevD.99.125006",
    journal = "Phys. Rev. D",
    volume = "99",
    number = "12",
    pages = "125006",
    year = "2019"
}

@book{Birrell:1982ix,
    author = "Birrell, N. D. and Davies, P. C. W.",
    title = "{Quantum Fields in Curved Space}",
    doi = "10.1017/CBO9780511622632",
    isbn = "978-0-511-62263-2, 978-0-521-27858-4",
    publisher = "Cambridge University Press",
    address = "Cambridge, UK",
    series = "Cambridge Monographs on Mathematical Physics",
    year = "1982"
}

@article{Hollands:2014eia,
    author = "Hollands, Stefan and Wald, Robert M.",
    title = "{Quantum fields in curved spacetime}",
    eprint = "1401.2026",
    archivePrefix = "arXiv",
    primaryClass = "gr-qc",
    doi = "10.1016/j.physrep.2015.02.001",
    journal = "Phys. Rept.",
    volume = "574",
    pages = "1--35",
    year = "2015"
}

@book{Wald:1995yp,
    author = "Wald, Robert M.",
    title = "{Quantum Field Theory in Curved Space-Time and Black Hole Thermodynamics}",
    isbn = "978-0-226-87027-4",
    publisher = "University of Chicago Press",
    address = "Chicago, IL",
    series = "Chicago Lectures in Physics",
    year = "1995"
}

@article{Cheng:2022xyr,
    author = "Cheng, Peng and Mao, Pujian",
    title = "{Soft theorems in curved spacetime}",
    eprint = "2206.11564",
    archivePrefix = "arXiv",
    primaryClass = "hep-th",
    doi = "10.1103/PhysRevD.106.L081702",
    journal = "Phys. Rev. D",
    volume = "106",
    number = "8",
    pages = "L081702",
    year = "2022"
}

@article{Mao:2023rca,
    author = "Mao, Pujian and Zhang, Kai-Yu",
    title = "{Soft theorems in de Sitter spacetime}",
    eprint = "2308.08861",
    archivePrefix = "arXiv",
    primaryClass = "hep-th",
    doi = "10.1007/JHEP01(2024)044",
    journal = "JHEP",
    volume = "01",
    pages = "044",
    year = "2024"
}

@article{Mao:2023dsy,
    author = "Mao, Pujian and Zhang, Kai-Yu and Zhou, Bochen",
    title = "{Near horizon linearized gravity and soft theorem}",
    eprint = "2311.03773",
    archivePrefix = "arXiv",
    primaryClass = "hep-th",
    doi = "10.1103/PhysRevD.109.065022",
    journal = "Phys. Rev. D",
    volume = "109",
    number = "6",
    pages = "065022",
    year = "2024"
}

@article{Mao:2024phf,
    author = "Mao, Pujian and Zhou, Bochen",
    title = "{Linearized gravity and soft graviton theorem in de Sitter spacetime}",
    eprint = "2405.12027",
    archivePrefix = "arXiv",
    primaryClass = "hep-th",
    doi = "10.1103/PhysRevD.110.065002",
    journal = "Phys. Rev. D",
    volume = "110",
    number = "6",
    pages = "065002",
    year = "2024"
}

@article{Hawking:2016msc,
    author = "Hawking, Stephen W. and Perry, Malcolm J. and Strominger, Andrew",
    title = "{Soft Hair on Black Holes}",
    eprint = "1601.00921",
    archivePrefix = "arXiv",
    primaryClass = "hep-th",
    doi = "10.1103/PhysRevLett.116.231301",
    journal = "Phys. Rev. Lett.",
    volume = "116",
    number = "23",
    pages = "231301",
    year = "2016"
}

@book{Strominger:2017zoo,
    author = "Strominger, Andrew",
    title = "{Lectures on the Infrared Structure of Gravity and Gauge Theory}",
    eprint = "1703.05448",
    archivePrefix = "arXiv",
    primaryClass = "hep-th",
    isbn = "978-0-691-17973-5",
    publisher = "Princeton University Press",
    year = "2018"
}

@article{Strominger:2013lka,
    author = "Strominger, Andrew",
    title = "{Asymptotic Symmetries of Yang-Mills Theory}",
    eprint = "1308.0589",
    archivePrefix = "arXiv",
    primaryClass = "hep-th",
    doi = "10.1007/JHEP07(2014)151",
    journal = "JHEP",
    volume = "07",
    pages = "151",
    year = "2014"
}

@article{Kapec:2014zla,
    author = "Kapec, Daniel and Lysov, Vyacheslav and Strominger, Andrew",
    title = "{Asymptotic Symmetries of Massless QED in Even Dimensions}",
    eprint = "1412.2763",
    archivePrefix = "arXiv",
    primaryClass = "hep-th",
    doi = "10.4310/ATMP.2017.v21.n7.a6",
    journal = "Adv. Theor. Math. Phys.",
    volume = "21",
    pages = "1747--1767",
    year = "2017"
}

@article{Campiglia:2015qka,
    author = "Campiglia, Miguel and Laddha, Alok",
    title = "{Asymptotic symmetries of QED and Weinberg{\textquoteright}s soft photon theorem}",
    eprint = "1505.05346",
    archivePrefix = "arXiv",
    primaryClass = "hep-th",
    doi = "10.1007/JHEP07(2015)115",
    journal = "JHEP",
    volume = "07",
    pages = "115",
    year = "2015"
}

@article{Kapec:2015ena,
    author = "Kapec, Daniel and Pate, Monica and Strominger, Andrew",
    title = "{New Symmetries of QED}",
    eprint = "1506.02906",
    archivePrefix = "arXiv",
    primaryClass = "hep-th",
    doi = "10.4310/ATMP.2017.v21.n7.a7",
    journal = "Adv. Theor. Math. Phys.",
    volume = "21",
    pages = "1769--1785",
    year = "2017"
}

@article{Lysov:2014csa,
    author = "Lysov, Vyacheslav and Pasterski, Sabrina and Strominger, Andrew",
    title = "{Low{\textquoteright}s Subleading Soft Theorem as a Symmetry of QED}",
    eprint = "1407.3814",
    archivePrefix = "arXiv",
    primaryClass = "hep-th",
    doi = "10.1103/PhysRevLett.113.111601",
    journal = "Phys. Rev. Lett.",
    volume = "113",
    number = "11",
    pages = "111601",
    year = "2014"
}

@article{Bosma:2023sxn,
    author = "Bosma, Jorrit and Geiller, Marc and Majumdar, Sucheta and Oblak, Blagoje",
    title = "{Radiative asymptotic symmetries of 3D Einstein-Maxwell theory}",
    eprint = "2311.09156",
    archivePrefix = "arXiv",
    primaryClass = "hep-th",
    doi = "10.21468/SciPostPhys.16.4.092",
    journal = "SciPost Phys.",
    volume = "16",
    number = "4",
    pages = "092",
    year = "2024"
}

@article{Shimizu:2025hfl,
    author = "Shimizu, Keito and Sugishita, Sotaro",
    title = "{Asymptotic symmetry and confinement in three-dimensional QED}",
    eprint = "2503.20173",
    archivePrefix = "arXiv",
    primaryClass = "hep-th",
    reportNumber = "KUNS-3041",
    doi = "10.1103/x92g-9w2h",
    journal = "Phys. Rev. D",
    volume = "112",
    number = "10",
    pages = "105001",
    year = "2025"
}

@article{Cotler:2024cia,
    author = "Cotler, Jordan and Jensen, Kristan and Prohazka, Stefan and Riegler, Max and Salzer, Jakob",
    title = "{Soft gravitons in three dimensions}",
    eprint = "2411.13633",
    archivePrefix = "arXiv",
    primaryClass = "hep-th",
    doi = "10.1007/JHEP07(2025)002",
    journal = "JHEP",
    volume = "07",
    pages = "002",
    year = "2025"
}

@article{He:2020ifr,
    author = "He, Temple and Mitra, Prahar",
    title = "{Covariant Phase Space and Soft Factorization in Non-Abelian Gauge Theories}",
    eprint = "2009.14334",
    archivePrefix = "arXiv",
    primaryClass = "hep-th",
    doi = "10.1007/JHEP03(2021)015",
    journal = "JHEP",
    volume = "03",
    pages = "015",
    year = "2021"
}

@article{Satishchandran:2019pyc,
    author = "Satishchandran, Gautam and Wald, Robert M.",
    title = "{Asymptotic behavior of massless fields and the memory effect}",
    eprint = "1901.05942",
    archivePrefix = "arXiv",
    primaryClass = "gr-qc",
    doi = "10.1103/PhysRevD.99.084007",
    journal = "Phys. Rev. D",
    volume = "99",
    number = "8",
    pages = "084007",
    year = "2019"
}

@article{Campoleoni:2019ptc,
    author = "Campoleoni, Andrea and Francia, Dario and Heissenberg, Carlo",
    title = "{Electromagnetic and color memory in even dimensions}",
    eprint = "1907.05187",
    archivePrefix = "arXiv",
    primaryClass = "hep-th",
    doi = "10.1103/PhysRevD.100.085015",
    journal = "Phys. Rev. D",
    volume = "100",
    number = "8",
    pages = "085015",
    year = "2019"
}

@article{He:2019pll,
    author = "He, Temple and Mitra, Prahar",
    title = "{Asymptotic symmetries in (d + 2)-dimensional gauge theories}",
    eprint = "1903.03607",
    archivePrefix = "arXiv",
    primaryClass = "hep-th",
    doi = "10.1007/JHEP10(2019)277",
    journal = "JHEP",
    volume = "10",
    pages = "277",
    year = "2019"
}

@article{Esmaeili:2019hom,
    author = "Esmaeili, Erfan",
    title = "{Asymptotic Symmetries of Maxwell Theory in Arbitrary Dimensions at Spatial Infinity}",
    eprint = "1902.02769",
    archivePrefix = "arXiv",
    primaryClass = "hep-th",
    doi = "10.1007/JHEP10(2019)224",
    journal = "JHEP",
    volume = "10",
    pages = "224",
    year = "2019"
}

@article{Freidel:2019ohg,
    author = {Freidel, Laurent and Hopfm{\"u}ller, Florian and Riello, Aldo},
    title = "{Asymptotic Renormalization in Flat Space: Symplectic Potential and Charges of Electromagnetism}",
    eprint = "1904.04384",
    archivePrefix = "arXiv",
    primaryClass = "hep-th",
    doi = "10.1007/JHEP10(2019)126",
    journal = "JHEP",
    volume = "10",
    pages = "126",
    year = "2019"
}

@article{He:2023bvv,
    author = "He, Temple and Mitra, Prahar",
    title = "{Asymptotic structure of higher dimensional Yang-Mills theory}",
    eprint = "2306.04571",
    archivePrefix = "arXiv",
    primaryClass = "hep-th",
    reportNumber = "CALT-TH 2023-015",
    doi = "10.21468/SciPostPhys.16.5.142",
    journal = "SciPost Phys.",
    volume = "16",
    number = "5",
    pages = "142",
    year = "2024"
}

@article{Weinberg:1965nx,
    author = "Weinberg, Steven",
    title = "{Infrared photons and gravitons}",
    doi = "10.1103/PhysRev.140.B516",
    journal = "Phys. Rev.",
    volume = "140",
    pages = "B516--B524",
    year = "1965"
}

@article{Strominger:2013jfa,
    author = "Strominger, Andrew",
    title = "{On BMS Invariance of Gravitational Scattering}",
    eprint = "1312.2229",
    archivePrefix = "arXiv",
    primaryClass = "hep-th",
    doi = "10.1007/JHEP07(2014)152",
    journal = "JHEP",
    volume = "07",
    pages = "152",
    year = "2014"
}

@article{He:2015zea,
    author = "He, Temple and Mitra, Prahar and Strominger, Andrew",
    title = "{2D Kac-Moody Symmetry of 4D Yang-Mills Theory}",
    eprint = "1503.02663",
    archivePrefix = "arXiv",
    primaryClass = "hep-th",
    doi = "10.1007/JHEP10(2016)137",
    journal = "JHEP",
    volume = "10",
    pages = "137",
    year = "2016"
}

@article{Strominger:2014pwa,
    author = "Strominger, Andrew and Zhiboedov, Alexander",
    title = "{Gravitational Memory, BMS Supertranslations and Soft Theorems}",
    eprint = "1411.5745",
    archivePrefix = "arXiv",
    primaryClass = "hep-th",
    doi = "10.1007/JHEP01(2016)086",
    journal = "JHEP",
    volume = "01",
    pages = "086",
    year = "2016"
}

@article{Pasterski:2015zua,
    author = "Pasterski, Sabrina",
    title = "{Asymptotic Symmetries and Electromagnetic Memory}",
    eprint = "1505.00716",
    archivePrefix = "arXiv",
    primaryClass = "hep-th",
    doi = "10.1007/JHEP09(2017)154",
    journal = "JHEP",
    volume = "09",
    pages = "154",
    year = "2017"
}

@article{Mao:2017axa,
    author = "Mao, Pujian and Ouyang, Hao and Wu, Jun-Bao and Wu, Xiaoning",
    title = "{New electromagnetic memories and soft photon theorems}",
    eprint = "1703.06588",
    archivePrefix = "arXiv",
    primaryClass = "hep-th",
    reportNumber = "CJQS-2017-003",
    doi = "10.1103/PhysRevD.95.125011",
    journal = "Phys. Rev. D",
    volume = "95",
    number = "12",
    pages = "125011",
    year = "2017"
}

@article{Pate:2017vwa,
    author = "Pate, Monica and Raclariu, Ana-Maria and Strominger, Andrew",
    title = "{Color Memory: A Yang-Mills Analog of Gravitational Wave Memory}",
    eprint = "1707.08016",
    archivePrefix = "arXiv",
    primaryClass = "hep-th",
    doi = "10.1103/PhysRevLett.119.261602",
    journal = "Phys. Rev. Lett.",
    volume = "119",
    number = "26",
    pages = "261602",
    year = "2017"
}

@article{Donnay:2015abr,
    author = "Donnay, Laura and Giribet, Gaston and Gonzalez, Hernan A. and Pino, Miguel",
    title = "{Supertranslations and Superrotations at the Black Hole Horizon}",
    eprint = "1511.08687",
    archivePrefix = "arXiv",
    primaryClass = "hep-th",
    doi = "10.1103/PhysRevLett.116.091101",
    journal = "Phys. Rev. Lett.",
    volume = "116",
    number = "9",
    pages = "091101",
    year = "2016"
}

@article{Donnay:2016ejv,
    author = "Donnay, Laura and Giribet, Gaston and Gonz{\'a}lez, Hern{\'a}n A. and Pino, Miguel",
    title = "{Extended Symmetries at the Black Hole Horizon}",
    eprint = "1607.05703",
    archivePrefix = "arXiv",
    primaryClass = "hep-th",
    doi = "10.1007/JHEP09(2016)100",
    journal = "JHEP",
    volume = "09",
    pages = "100",
    year = "2016"
}

@article{Mao:2016pwq,
    author = "Mao, Pujian and Wu, Xiaoning and Zhang, Hongbao",
    title = "{Soft hairs on isolated horizon implanted by electromagnetic fields}",
    eprint = "1606.03226",
    archivePrefix = "arXiv",
    primaryClass = "hep-th",
    doi = "10.1088/1361-6382/aa59da",
    journal = "Class. Quant. Grav.",
    volume = "34",
    number = "5",
    pages = "055003",
    year = "2017"
}

@article{Hawking:2016sgy,
    author = "Hawking, Stephen W. and Perry, Malcolm J. and Strominger, Andrew",
    title = "{Superrotation Charge and Supertranslation Hair on Black Holes}",
    eprint = "1611.09175",
    archivePrefix = "arXiv",
    primaryClass = "hep-th",
    doi = "10.1007/JHEP05(2017)161",
    journal = "JHEP",
    volume = "05",
    pages = "161",
    year = "2017"
}

@article{Donnay:2018ckb,
    author = "Donnay, Laura and Giribet, Gaston and Gonz{\'a}lez, Hern{\'a}n A. and Puhm, Andrea",
    title = "{Black hole memory effect}",
    eprint = "1809.07266",
    archivePrefix = "arXiv",
    primaryClass = "hep-th",
    reportNumber = "CPHT-RR021.042018",
    doi = "10.1103/PhysRevD.98.124016",
    journal = "Phys. Rev. D",
    volume = "98",
    number = "12",
    pages = "124016",
    year = "2018"
}

@article{Rahman:2019bmk,
    author = "Rahman, Adel A. and Wald, Robert M.",
    title = "{Black Hole Memory}",
    eprint = "1912.12806",
    archivePrefix = "arXiv",
    primaryClass = "gr-qc",
    doi = "10.1103/PhysRevD.101.124010",
    journal = "Phys. Rev. D",
    volume = "101",
    number = "12",
    pages = "124010",
    year = "2020"
}

@article{Grumiller:2019fmp,
    author = "Grumiller, Daniel and P{\'e}rez, Alfredo and Sheikh-Jabbari, M. M. and Troncoso, Ricardo and Zwikel, C{\'e}line",
    title = "{Spacetime structure near generic horizons and soft hair}",
    eprint = "1908.09833",
    archivePrefix = "arXiv",
    primaryClass = "hep-th",
    reportNumber = "Preprint: CECS-PHY-18/01, IPM/P-2019/009, TUW-18-03",
    doi = "10.1103/PhysRevLett.124.041601",
    journal = "Phys. Rev. Lett.",
    volume = "124",
    number = "4",
    pages = "041601",
    year = "2020"
}

@article{Cheng:2022xgm,
    author = "Cheng, Peng and Mao, Pujian",
    title = "{Soft gluon theorems in curved spacetime}",
    eprint = "2211.00031",
    archivePrefix = "arXiv",
    primaryClass = "hep-th",
    doi = "10.1103/PhysRevD.107.065010",
    journal = "Phys. Rev. D",
    volume = "107",
    number = "6",
    pages = "065010",
    year = "2023"
}

@article{Low:1958sn,
    author = "Low, F. E.",
    title = "{Bremsstrahlung of very low-energy quanta in elementary particle collisions}",
    doi = "10.1103/PhysRev.110.974",
    journal = "Phys. Rev.",
    volume = "110",
    pages = "974--977",
    year = "1958"
}

@article{Burnett:1967km,
    author = "Burnett, T. H. and Kroll, Norman M.",
    title = "{Extension of the low soft photon theorem}",
    doi = "10.1103/PhysRevLett.20.86",
    journal = "Phys. Rev. Lett.",
    volume = "20",
    pages = "86",
    year = "1968"
}

@article{Casali:2014xpa,
    author = "Casali, Eduardo",
    title = "{Soft sub-leading divergences in Yang-Mills amplitudes}",
    eprint = "1404.5551",
    archivePrefix = "arXiv",
    primaryClass = "hep-th",
    doi = "10.1007/JHEP08(2014)077",
    journal = "JHEP",
    volume = "08",
    pages = "077",
    year = "2014"
}

@article{Schwab:2014xua,
    author = "Schwab, Burkhard U. W. and Volovich, Anastasia",
    title = "{Subleading Soft Theorem in Arbitrary Dimensions from Scattering Equations}",
    eprint = "1404.7749",
    archivePrefix = "arXiv",
    primaryClass = "hep-th",
    doi = "10.1103/PhysRevLett.113.101601",
    journal = "Phys. Rev. Lett.",
    volume = "113",
    number = "10",
    pages = "101601",
    year = "2014"
}

@article{Campiglia:2014yka,
    author = "Campiglia, Miguel and Laddha, Alok",
    title = "{Asymptotic symmetries and subleading soft graviton theorem}",
    eprint = "1408.2228",
    archivePrefix = "arXiv",
    primaryClass = "hep-th",
    doi = "10.1103/PhysRevD.90.124028",
    journal = "Phys. Rev. D",
    volume = "90",
    number = "12",
    pages = "124028",
    year = "2014"
}

@article{Zlotnikov:2014sva,
    author = "Zlotnikov, Michael",
    title = "{Sub-sub-leading soft-graviton theorem in arbitrary dimension}",
    eprint = "1407.5936",
    archivePrefix = "arXiv",
    primaryClass = "hep-th",
    doi = "10.1007/JHEP10(2014)148",
    journal = "JHEP",
    volume = "10",
    pages = "148",
    year = "2014"
}

@article{Kalousios:2014uva,
    author = "Kalousios, Chrysostomos and Rojas, Francisco",
    title = "{Next to subleading soft-graviton theorem in arbitrary dimensions}",
    eprint = "1407.5982",
    archivePrefix = "arXiv",
    primaryClass = "hep-th",
    reportNumber = "ICTP-SAIFR-2014-005",
    doi = "10.1007/JHEP01(2015)107",
    journal = "JHEP",
    volume = "01",
    pages = "107",
    year = "2015"
}

@article{Campiglia:2016jdj,
    author = "Campiglia, Miguel and Laddha, Alok",
    title = "{Sub-subleading soft gravitons: New symmetries of quantum gravity?}",
    eprint = "1605.09094",
    archivePrefix = "arXiv",
    primaryClass = "gr-qc",
    doi = "10.1016/j.physletb.2016.11.046",
    journal = "Phys. Lett. B",
    volume = "764",
    pages = "218--221",
    year = "2017"
}

@article{Conde:2016rom,
    author = "Conde, Eduardo and Mao, Pujian",
    title = "{BMS Supertranslations and Not So Soft Gravitons}",
    eprint = "1612.08294",
    archivePrefix = "arXiv",
    primaryClass = "hep-th",
    doi = "10.1007/JHEP05(2017)060",
    journal = "JHEP",
    volume = "05",
    pages = "060",
    year = "2017"
}

@article{Laddha:2017ygw,
    author = "Laddha, Alok and Sen, Ashoke",
    title = "{Sub-subleading Soft Graviton Theorem in Generic Theories of Quantum Gravity}",
    eprint = "1706.00759",
    archivePrefix = "arXiv",
    primaryClass = "hep-th",
    doi = "10.1007/JHEP10(2017)065",
    journal = "JHEP",
    volume = "10",
    pages = "065",
    year = "2017"
}

@article{Averin:2016ybl,
    author = "Averin, Artem and Dvali, Gia and Gomez, Cesar and Lust, Dieter",
    title = "{Gravitational Black Hole Hair from Event Horizon Supertranslations}",
    eprint = "1601.03725",
    archivePrefix = "arXiv",
    primaryClass = "hep-th",
    reportNumber = "LMU--ASC-05-16, MPP--2016--3",
    doi = "10.1007/JHEP06(2016)088",
    journal = "JHEP",
    volume = "06",
    pages = "088",
    year = "2016"
}

@article{Shi:2016jtn,
    author = "Shi, Changfu and Mei, Jianwei",
    title = "{Extended Symmetries at Black Hole Horizons in Generic Dimensions}",
    eprint = "1611.09491",
    archivePrefix = "arXiv",
    primaryClass = "gr-qc",
    doi = "10.1103/PhysRevD.95.104053",
    journal = "Phys. Rev. D",
    volume = "95",
    number = "10",
    pages = "104053",
    year = "2017"
}

@article{Choi:2018oel,
    author = "Choi, Sangmin and Akhoury, Ratindranath",
    title = "{Soft Photon Hair on Schwarzschild Horizon from a Wilson Line Perspective}",
    eprint = "1809.03467",
    archivePrefix = "arXiv",
    primaryClass = "hep-th",
    doi = "10.1007/JHEP12(2018)074",
    journal = "JHEP",
    volume = "12",
    pages = "074",
    year = "2018"
}

@article{Choi:2019fuq,
    author = "Choi, Sangmin and Sandeep Pradhan, S. and Akhoury, Ratindranath",
    title = "{Supertranslation Hair of Schwarzschild Black Hole: A Wilson Line Perspective}",
    eprint = "1910.05882",
    archivePrefix = "arXiv",
    primaryClass = "hep-th",
    doi = "10.1007/JHEP01(2020)013",
    journal = "JHEP",
    volume = "01",
    pages = "013",
    year = "2020"
}

@article{Adami:2020amw,
    author = "Adami, H. and Grumiller, D. and Sadeghian, S. and Sheikh-Jabbari, M. M. and Zwikel, C.",
    title = "{T-Witts from the horizon}",
    eprint = "2002.08346",
    archivePrefix = "arXiv",
    primaryClass = "hep-th",
    reportNumber = "IPM/P-2019/041, TUW-20-02",
    doi = "10.1007/JHEP04(2020)128",
    journal = "JHEP",
    volume = "04",
    pages = "128",
    year = "2020"
}

@article{Liu:2022uox,
    author = "Liu, Hai-Shan and Mao, Pujian",
    title = "{Near horizon gravitational charges}",
    eprint = "2201.10308",
    archivePrefix = "arXiv",
    primaryClass = "hep-th",
    doi = "10.1007/JHEP05(2022)123",
    journal = "JHEP",
    volume = "05",
    pages = "123",
    year = "2022"
}

@article{Zhuge:2026oap,
    author = "Zhuge, Jin-Peng and Cheng, Peng",
    title = "{Near-horizon soft theorems from local Ward identities}",
    eprint = "2609.39288",
    archivePrefix = "arXiv",
    primaryClass = "hep-th",
    month = "9",
    year = "2026"
}

@article{Pasterski:2016qvg,
    author = "Pasterski, Sabrina and Shao, Shu-Heng and Strominger, Andrew",
    title = "{Flat Space Amplitudes and Conformal Symmetry of the Celestial Sphere}",
    eprint = "1701.00049",
    archivePrefix = "arXiv",
    primaryClass = "hep-th",
    doi = "10.1103/PhysRevD.96.065026",
    journal = "Phys. Rev. D",
    volume = "96",
    number = "6",
    pages = "065026",
    year = "2017"
}

@article{Pasterski:2017kqt,
    author = "Pasterski, Sabrina and Shao, Shu-Heng",
    title = "{Conformal basis for flat space amplitudes}",
    eprint = "1705.01027",
    archivePrefix = "arXiv",
    primaryClass = "hep-th",
    doi = "10.1103/PhysRevD.96.065022",
    journal = "Phys. Rev. D",
    volume = "96",
    number = "6",
    pages = "065022",
    year = "2017"
}

@article{Raclariu:2021zjz,
    author = "Raclariu, Ana-Maria",
    title = "{Lectures on Celestial Holography}",
    eprint = "2107.02075",
    archivePrefix = "arXiv",
    primaryClass = "hep-th",
    month = "7",
    year = "2021"
}

@article{Donnay:2023mrd,
    author = "Donnay, Laura",
    title = "{Celestial holography: An asymptotic symmetry perspective}",
    eprint = "2310.12922",
    archivePrefix = "arXiv",
    primaryClass = "hep-th",
    doi = "10.1016/j.physrep.2024.04.003",
    journal = "Phys. Rept.",
    volume = "1073",
    pages = "1--41",
    year = "2024"
}

@article{Hollands:2003ie,
    author = "Hollands, Stefan and Ishibashi, Akihiro",
    title = "{Asymptotic flatness and Bondi energy in higher dimensional gravity}",
    eprint = "gr-qc/0304054",
    archivePrefix = "arXiv",
    doi = "10.1063/1.1829152",
    journal = "J. Math. Phys.",
    volume = "46",
    pages = "022503",
    year = "2005"
}

@article{Hollands:2004ac,
    author = "Hollands, Stefan and Wald, Robert M.",
    title = "{Conformal null infinity does not exist for radiating solutions in odd spacetime dimensions}",
    eprint = "gr-qc/0407014",
    archivePrefix = "arXiv",
    doi = "10.1088/0264-9381/21/22/008",
    journal = "Class. Quant. Grav.",
    volume = "21",
    pages = "5139--5146",
    year = "2004"
}

@article{Tanabe:2011es,
    author = "Tanabe, Kentaro and Kinoshita, Shunichiro and Shiromizu, Tetsuya",
    title = "{Asymptotic flatness at null infinity in arbitrary dimensions}",
    eprint = "1104.0303",
    archivePrefix = "arXiv",
    primaryClass = "gr-qc",
    reportNumber = "YITP-11-40",
    doi = "10.1103/PhysRevD.84.044055",
    journal = "Phys. Rev. D",
    volume = "84",
    pages = "044055",
    year = "2011"
}

@article{Crispino:2000jx,
    author = "Crispino, Luis C. B. and Higuchi, Atsushi and Matsas, George E. A.",
    title = "{Quantization of the electromagnetic field outside static black holes and its application to low-energy phenomena}",
    eprint = "gr-qc/0011070",
    archivePrefix = "arXiv",
    doi = "10.1103/PhysRevD.80.029906",
    journal = "Phys. Rev. D",
    volume = "63",
    pages = "124008",
    year = "2001",
    note = "[Erratum: Phys.Rev.D 80, 029906 (2009)]"
}

@article{Lopez-Ortega:2006vjp,
    author = "Lopez-Ortega, A.",
    title = "{Electromagnetic quasinormal modes of D-dimensional black holes}",
    eprint = "gr-qc/0605034",
    archivePrefix = "arXiv",
    doi = "10.1007/s10714-006-0358-2",
    journal = "Gen. Rel. Grav.",
    volume = "38",
    pages = "1747--1770",
    year = "2006"
}

@article{Chrusciel:2012jk,
    author = "Chrusciel, Piotr T. and Lopes Costa, Joao and Heusler, Markus",
    title = "{Stationary Black Holes: Uniqueness and Beyond}",
    eprint = "1205.6112",
    archivePrefix = "arXiv",
    primaryClass = "gr-qc",
    doi = "10.12942/lrr-2012-7",
    journal = "Living Rev. Rel.",
    volume = "15",
    pages = "7",
    year = "2012"
}

@article{Conde:2016csj,
    author = "Conde, Eduardo and Mao, Pujian",
    title = "{Remarks on asymptotic symmetries and the subleading soft photon theorem}",
    eprint = "1605.09731",
    archivePrefix = "arXiv",
    primaryClass = "hep-th",
    doi = "10.1103/PhysRevD.95.021701",
    journal = "Phys. Rev. D",
    volume = "95",
    number = "2",
    pages = "021701",
    year = "2017"
}
\end{document}